\documentclass[journal,comsoc,twocolumn]{IEEEtran}
\pdfoutput=1
\usepackage{orgnizations/journalPackages}

\begin{document}
\title{A design framework for all-digital mmWave massive MIMO with per-antenna nonlinearities} 
\author{Mohammed~Abdelghany, Ali A. Farid, %
 Upamanyu Madhow,~\IEEEmembership{Fellow,~IEEE}, and Mark J. W. Rodwell,~\IEEEmembership{Fellow,~IEEE}.%
\thanks{M. Abdelghany A. Farid, U. Madhow, and M. Rodwell are with the Department of Electrical and Computer Engineering,
University of California at Santa Barbara, Santa Barbara, CA 93106 USA
(e-mail: mabdelghany@ucsb.edu; afarid@ece.ucsb.edu; madhow@ece.ucsb.edu; rodwell@ucsb.edu).}}
\markboth{\tiny This work has been submitted to the IEEE for possible publication. Copyright may be transferred without notice, after which this version may no longer be accessible.}%
{Shell \MakeLowercase{\textit{et al.}}: Bare Demo of IEEEtran.cls for Journals}
\maketitle
\begin{abstract}

Millimeter wave MIMO combines the benefits of compact antenna arrays with a large number of elements and massive bandwidths, so that fully digital beamforming 
has the potential of supporting a large number of simultaneous users with {\it per user} data rates of multiple gigabits/sec (Gbps). 
In this paper, we develop an analytical model for the impact of nonlinearities in such a system, and illustrate its utility in providing 
hardware design guidelines regarding two key challenges: the low available precision of analog-to-digital conversion at high sampling rates, and 
nonlinearities in ultra-high speed radio frequency (RF) and baseband circuits.  We consider linear minimum mean square error (LMMSE) reception for a
multiuser MIMO uplink, and provide performance guarantees based on two key concepts: (a) summarization of the impact of per-antenna nonlinearities via a quantity that we term
the ``intrinsic SNR'', (b) using linear MMSE performance in an ideal system without nonlinearities to bound that in our non-ideal system. 
For our numerical results, we employ nominal parameters corresponding to outdoor picocells operating at a carrier frequency of 140 GHz, with a data rate of 10 Gbps per user.

\end{abstract}
\section{Introduction}\label{Intro}

We present an analytical framework for quantifying the impact of nonlinearities on millimeter wave (mmWave) multiuser MIMO.
Most recent research on mmWave communication has focused on radio frequency (RF) beamforming, which supports a single user at a time, or hybrid beamforming, where the number of supported users equals the number of RF chains, typically set to be much smaller than the number of antennas. However, recent advances in silicon realizations of mmWave hardware imply that scaling the number of RF chains with the number of antennas is on the cusp of feasibility, which opens up the possibility of fully digital spatial processing. This implies that multiuser detection can be employed to support a large number of simultaneous users, since the small carrier wavelengths at mmWave bands enable the realization of compact antenna arrays with a large number of elements. Furthermore, the massive available bandwidths imply that {\it per-user}
data rates of multiple gigabits/second (Gbps) can be supported in such a system.  

The running example for our numerical results is a 140~GHz picocellular uplink, with a linear array with 256 elements supporting up to 128 simultaneous users at a range of up to 100~m, using linear minimum mean square error (LMMSE) reception.  Using a symbol rate of 5 Gbaud and QPSK modulation provides per user data rates of 10 Gbps, resulting in an aggregate throughput of up to 1.28 Tbps! As we shall show, in the presence of nonlinearities, it is advantageous to operate at a smaller load factor (defined as the ratio of the number of simultaneous users to the number of antennas). However, a very large aggregate throughput of 160 Gbps is obtained even 
when the load factor is reduced to $\frac{1}{16}$.

Besides the enormous aggregate throughput, from a hardware perspective, the all-digital solution is more efficient in terms of power and area compared to the hybrid architecture \cite{yan2018performance}. Nevertheless, nonlinearities present a fundamental challenge in realizing the envisioned system.  Wideband RF and baseband circuits scaled via relatively low-end silicon (e.g., CMOS) semiconductor processes exhibit significant nonlinearities, while the analog-to-digital converters (ADCs) available at multi-GHz sampling rates have relatively low precision.  Our goal in this paper is to provide a framework that enables designers to determine the permissible levels of nonlinearities for providing desired system-level performance guarantees.  

\subsection{Contributions} 

Our analytical framework is based on two core concepts:\\
(a) We show that the impact of per-antenna nonlinearities is effectively summarized by a quantity that we term the {\it intrinsic SNR,} corresponding to a normalized version
of the nonlinearity. Key elements of this characterization are a Bussgang decomposition and the observation that, even for a moderate number of simultaneous users and without rich scattering, the antenna input is well modeled as zero-mean complex Gaussian. We show that the matched filter bound on the effective SNR for a given user,
which captures the effect of the self-noise generated by per-antenna nonlinearities, depends only on four parameters: the user's SNR, the intrinsic SNR, the load factor and 
a power control factor which summarizes the variations in received signal power across users.\\
(b) We show that a pessimistic estimate of the degradation in performance due to multiuser interference can be obtained by analyzing (theoretically and/or numerically) an {\it ideal} system without nonlinearities.  Thus, we can provide a lower bound on the output signal-to-interference-plus-noise ratio (SINR) of a linear MMSE receiver, 
accounting for both nonlinearities and multiuser interference.

Combining these two concepts, averaging over the spatial distribution of users, and specializing to an edge user in the cell, allows us to provide analytical guidelines for maximum permissible levels of nonlinearities in order to provide a desired system-level performance guarantees (e.g., on outage probabilities).
We consider third order RF and baseband nonlinearities that can be specified using the so-called 1~dB compression point \cite{razavi1998rf}, termed
$P_{\text{1dB}}$.  The per-antenna ADCs for the in-phase and quadrature components are modeled as overloaded uniform quantizers optimized (for a specified number of bits) to minimize the mean square error with zero mean Gaussian input. Using our framework, we are able to provide compact design prescriptions for $P_{\text{1dB}}$ and the number of ADC bits. For example, for a load factor of 1/2, the system can work with 4-bit ADC and passband/baseband $P_{\text{1dB}}$ of 8.4 dB / 5 dB. On the other hand, 2-bit ADC with passband/baseband $P_{\text{1dB}}$ of 1.4 dB / -1 dB suffice to work properly with a load factor of 1/16. We present extensive simulations verifying our analytical predictions and prescriptions.

\subsection{Related Work} 

While the focus in the present paper is on mmWave massive MIMO, there is a significant body of closely related recent research on the effect of nonlinearities
on multiuser massive MIMO at lower carrier frequencies.  Most of this prior work also employs Bussgang's theorem \cite{bussgang} to model the 
effect of nonlinearities, both for uplink reception and downlink precoding.  Our discussion here is limited to the literature on uplink massive MIMO,
since that is the focus of the present paper, but the design framework for modeling downlink nonlinearities such as power amplifiers and digital-to-analog converters (DACs)
is well known to be entirely analogous.

The line of sight (LoS) channel model used in our performance evaluation is different from that in much of this prior work, which employs
models that are better matched to the propagation environments at lower frequencies.  However, our analytical framework is quite general, and
can be used to obtain design prescriptions for lower carrier frequencies as well.  Conversely, many of the general observations emerging from prior work at lower carrier
frequencies are consistent with the conclusions in the present paper, given a common underlying mathematical framework that employs the Bussgang decomposition and
exploits the relaxation of hardware constraints enabled by the increase in the number of antennas. In the following, we briefly review this prior work in order to place the
contributions of the present paper in perspective.  


The potential for relaxing hardware constraints by increasing the number of antennas is clearly brought out by the theoretical
results in \cite{bjornson2014massive}, which show that the performance degradation due to hardware impairments vanishes asymptotically
as the number of base station antennas gets large.  The same trend holds for a finite but large number of antennas, as is clear from the results in  \cite{fan2015uplink,z2016spectral,xu2019uplink}, which study the spectral efficiency of quantized massive MIMO over frequency nonselective Rayleigh and Rician fading channels using maximum ratio combining. Another interesting conclusion from the simulations of  \cite{z2016spectral} is that, for Rician fading, the system is more vulnerable to drastic quantization as the relative strength of the specular component increases.  Thus, the LoS model considered in this paper may be a worst-case scenario
for obtaining design prescriptions regarding nonlinearities.

The impact of imperfect power control for quantized massive MIMO over frequency nonselective channels is included in the analysis in \cite{mollen2017achievable, jacobsson2017throughput}.  Using spectral efficiency as a performance measure, an example conclusion from \cite{mollen2017achievable} is that 3-bit ADC suffices for a system with 100 antennas serving 10 users at a spectral efficiency of 3.5 bits per channel use, with 4-bit ADCs recommended to handle imperfections in power control
and automatic gain control.  Similar conclusions are obtained in \cite{jacobsson2017throughput}, which shows moderate drops in spectral efficiency due
to imperfect power control. 

The impact of quantization on multiuser OFDM MIMO over a frequency-selective channel 
is studied in  \cite{studer2016quantized}, with a focus on low-complexity channel estimation and data detection.
The simulations in this paper show that, for the models considered, 4-bit ADC is sufficient to achieve a near-optimal performance (in terms of packet error rate) for a load factor of 1/8 or lower.  More recent work with a similar model \cite{j2018massive} employs a Bussgang-based analysis for the joint distortion introduced by nonlinear low-noise amplifiers, phase noise, and finite-resolution ADCs, and demonstrates its accuracy by comparing analytical predictions with simulations.

In comparison with the existing literature, the key conceptual novelty in the present paper is that we provide an analytical framework for mapping {\it system-level} performance goals to {\it hardware design} prescriptions for per-antenna nonlinearities.  The theoretical foundation for this mapping is our observation (Theorem V.2) that
an ideal system without nonlinearities provides a means of obtaining pessimistic performance estimates, together with our abstraction of self-noise via 
intrinsic SNR and the associated matched filter bound (Theorem V.1).
Thus, while prior work such as  \cite{studer2016quantized, j2018massive} demonstrates
the accuracy of Bussgang modeling and assesses design tradeoffs in particular scenarios, we are able to provide a general framework which provides {\it compact}
prescriptions that hardware designers can apply to design RF chains jointly with ADCs, by considering the cascade of passband amplifiers, baseband amplifiers and ADCs as the nonlinearities employed in our performance evaluation. Finally, unlike prior work on fading channels, we 
employ a LoS model which is a more suitable abstraction for mmWave channels \cite{LoS1,LoS2,LoS3,LoS4}. 

A preliminary version of this work has appeared in a conference paper \cite{AsilomarPaper}. In this paper, we provide a comprehensive analysis,
including proofs that were omitted in \cite{AsilomarPaper}, along with a more extensive set of numerical results.  We also study the impact of power control on our system-level performance objectives.  The system model is also different in some details from \cite{AsilomarPaper} in order to more closely model the hardware
designs that we are currently engaged in: we now include the impact of baseband as well as RF nonlinearities, and 
consider a more reasonable field of view for the base station array.

\section{System Model}\label{systemmodel}

\begin{figure}
\centering
  \includegraphics[trim=0cm 0cm 0cm 0cm,clip,scale=0.7]{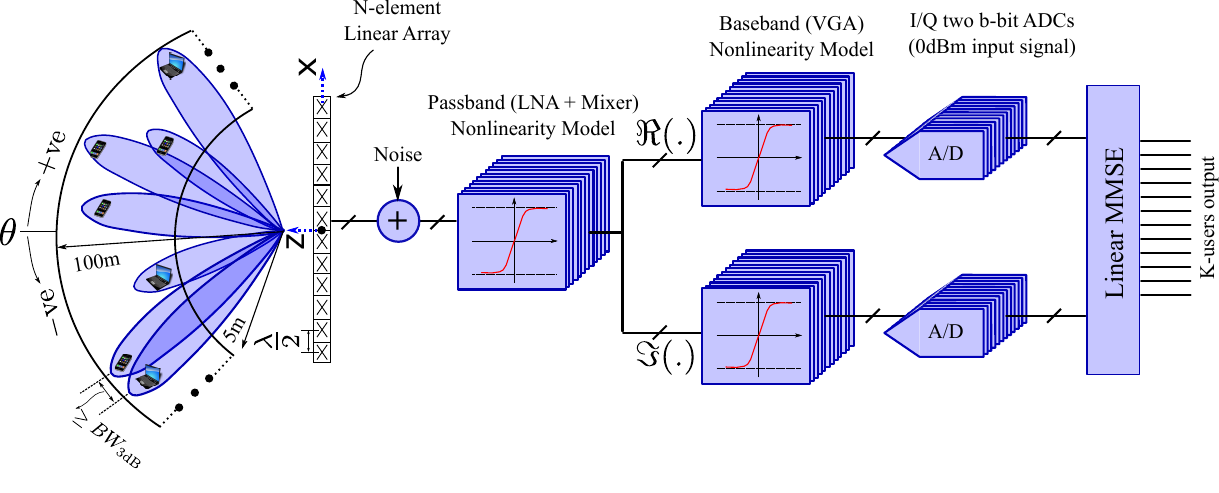}
  \caption{The cell size is constrained radially between $R_{\text{min}}$ and $R_{\text{max}}$ and angularly between $-\pi/3 \leq\theta \leq \pi/3$. $\text{BW}_{3 \text{dB}}$ stands for the 3 dB beamwidth. The passband and baseband nonlinearities are modeled by saturated third order polynomials. An overloaded uniform ADC with $b$ bits per dimension, optimized for a zero-mean standard Gaussian random variable, is used. Linear MMSE reception is employed after digitization.}
    \label{SystemModel}
\end{figure}

Fig. \ref{SystemModel} shows the system model. 
The base station performs horizontal scanning with a 1D half-wavelength spaced $N$-element array.
Let $K$ denotes the number of simultaneous users, and $\beta = \frac{K}{N}$ the {\it load factor.} 

We assume a line-of-sight (LoS) channel between the base station and each mobile.  The direction of arrival (DoA) from the $k^{\text{th}}$ mobile
is denoted by $\theta_k$, and corresponds to spatial frequency  $\Omega_k=2\pi\frac{d_x}{\lambda}\sin{\theta_k}$, where $\lambda$ denotes the carrier wavelength
and $d_x$ denotes the inter-element spacing, set to $\frac{\lambda}{2}$ in our numerical results. The $N \times 1$ spatial channel for mobile $k$ is given by
\begin{align} \label{LoS_channel}
    \mathbf{h}_k =A_k e^{j\phi_k}\, [1 \, e^{j\Omega_{k}} \, e^{j2\Omega_{k}} \, \hdots \, e^{j(N-1)\Omega_{k}} ]^\intercal,
\end{align}
where $\phi_k$ is an arbitrary phase shift and $A_k^2=\left(\frac{ \lambda}{4\pi R_k}\right)^{\mathclap{2}}$ depends on the radial location $R_k$ of mobile $k$, using the Friis formula for path loss.
Each mobile is assumed to be able to perform ideal transmit beamforming towards the base station.

The cascade of the nonlinearities described in Sections \ref{sec:nonlinearity_model} and \ref{sec:ADC_model}
is modeled as a complex baseband equivalent nonlinearity $g( \cdot )$.  
The complex baseband received signal vector $\mathbf{z}$ at the base station is therefore given by
\begin{align} \label{recd}
    \mathbf{z}=g(\mathbf{y})= g \left( \underbrace{\mathbf{H}}_{N\times K}\mathbf{x} + \mathbf{n} \right),
\end{align}
where $\mathbf{H}=[\mathbf{h}_1 \mathbf{h}_2 \hdots \mathbf{h}_K]$ is the channel matrix, $\mathbf{x} = [x_1,...,x_K]^T$ is the vector of symbols (normalized to unit energy: $\mathbb{E} \left[ |{x}_k |^2 \right]=1$) transmitted by the mobiles,  ${\mathbf{n}} \sim \mathcal{CN}(\mathbf{0},\sigma_{n}^2\mathbf{I})$ is the thermal AWGN vector, and $g( \cdot )$ is the effective per-antenna nonlinearity in complex baseband.


We note that the linear MMSE receiver used in the digital backend accounts for the self-noise due to nonlinearities (characterized in a later section), as well as interference and thermal noise.

\noindent
{\bf Running example:} We provide the link budget analysis for the envisioned  system in Appendix \ref{Appendix0}. We assume $N=256$ antennas, and load factor $\beta$ ranging from $\frac{1}{16}$ to $\frac{1}{2}$ (i.e., $K$ ranging from $16$ to $128$). We assume 5 Gbaud symbol rate, with each user employing QPSK modulation.  Ignoring channel coding overhead, the data rate per user is 10 Gbps, and the aggregate throughput ranges from 0.16 to 1.28 Tbps.

In the remainder of this section, we characterize the statistics of the received signal at each antenna and describe the nonlinearities modeled considered in our numerical results.  

\subsection{Per-antenna Received Signal Statistics} 

\begin{figure}
\centering
 \includegraphics[trim=7.5cm 0 7.5cm 0,clip,scale=0.4]{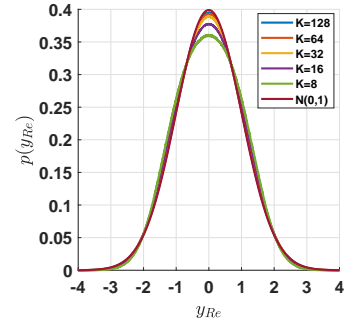}
\caption{The pdf of the standard normal distribution and the histogram of the normalized real/imaginary part of the received signal  at each antenna element when K users transmit. }
    \label{SignalStat}
\end{figure}

The input to the effective complex baseband nonlinearity $g ( \cdot )$ at, say, antenna $m$, is given by
\begin{equation} \label{nonlinearity_input}
y_m  = \sum_{k=1}^K A_k e^{j \phi_k} x_k  e^{j m \Omega_k}.
\end{equation}
For a uniform spatial distribution of users over the region of interest, the amplitudes $\{ A_k \}$ and spatial frequencies $\{ \Omega_k \}$
are independent and identically distributed (i.i.d.).  The phases $\{ \phi_k \}$ are uniform over $[0, 2 \pi ]$, and $x_k$ are i.i.d. QPSK symbols. 
By virtue of the central limit theorem (CLT), the received signal is well modeled as zero-mean complex Gaussian for large $K$, and jointly Gaussian across antennas. We have verified empirically, histogram comparisons, quantile-quantile plots and KL divergence computations, that this Gaussian approximation holds for even moderate number of mobiles (e.g., $K=8$) in all settings that we have considered. Fig.~\ref{SignalStat}~(a) illustrates a comparison between the histogram of the normalized real/imaginary component of the received signal and the standard normal distribution $\mathcal{N}(0,1)$.  

In terms of technical conditions for applying the CLT, we note that, with no power control, finite variance for each term follows from 
enforcing a lower bound on the distance from mobile to base station, which leads to an upper bound on $A_k$ for our free space propagation model.
We also note that, if the $A_k$ become dependent due to power control, then the CLT can be applied conditioned on $\{ A_k \}$.


\subsection{Passband and Baseband Nonlinearity Model}  \label{sec:nonlinearity_model}

 \begin{figure}
\centering
  \includegraphics[trim=0cm 0cm 0cm 0cm,clip,scale=0.7]{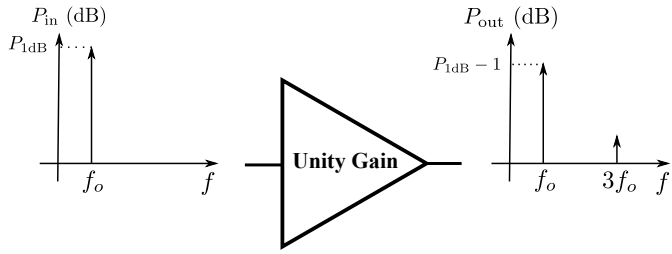}
  \caption{The 1~dB compression point ($P_{\text{1dB}}$) is defined as the input power at which the output power of the desired
  sinusoid (at $f_o$) is compressed by 1~dB.}
    \label{P1dBExplained}
\end{figure}
 
The passband nonlinearity arises in the low noise amplifier and the mixer, while the baseband nonlinearity 
is in the variable gain amplifier. We model each nonlinearity as a saturated third-order polynomial function with a nominal gain of unity.  The function is parametrized by the 1~dB compression point ($P_{\text{1dB}}$) \cite{razavi1998rf}, defined as the input power of a sinusoid of frequency $f_o$ (taken to be the carrier frequency) at which the output power
is reduced by 1~dB relative to the nominal. The concept is illustrated in Fig. \ref{P1dBExplained}.  

The gain compression for the passband nonlinearity depends on the absolute value of the complex baseband signal, while the gain compression depends on the absolute value of the I and Q components for the baseband nonlinearity. 

The third-order nonlinearity is written in terms of $P_{\text{1dB}}$ as follows:
\begin{align}
g(y(t))=\begin{cases}
y(t)(1-\frac{0.44|y(t)|^2}{3 {P}_{1\text{dB}}}) & \text{ if } |y(t)|^2\leq \frac{{P}_{1\text{dB}}}{0.44} \\ 
 \frac{y(t)}{|y(t)|}\sqrt{{{{P}_{1\text{dB}}}}} & \text{ if } |y(t)|^2> \frac{{P}_{1\text{dB}}}{0.44}
\end{cases}.
 \end{align}
Fig. \ref{Nonlinearity} (a) illustrates the distribution of the input powers of the passband and baseband nonlinearities, along with example 
input/output (I/O) characteristics. In this work, we consider the nonlinearities to be memoryless and free of phase distortion.

\subsection{ADC Model} \label{sec:ADC_model}

\begin{figure}
\centering
\subfloat[Third-order nonlinearities]{ \includegraphics[trim=11cm 0 3.5cm 0,clip,scale=0.33]{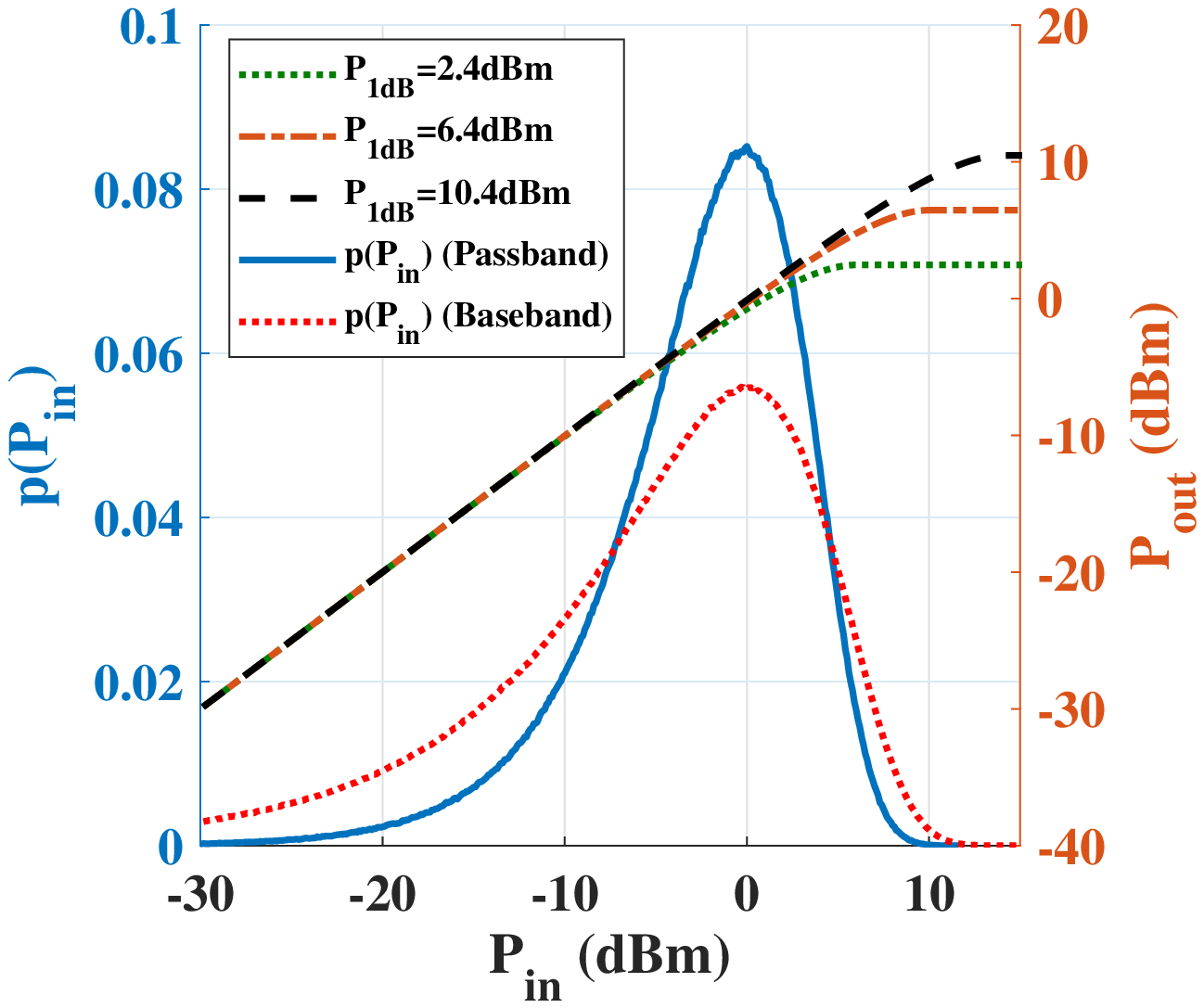}}
\subfloat[Overloaded uniform ADC]{ \includegraphics[trim=7cm 0 7cm 0,clip,scale=0.33]{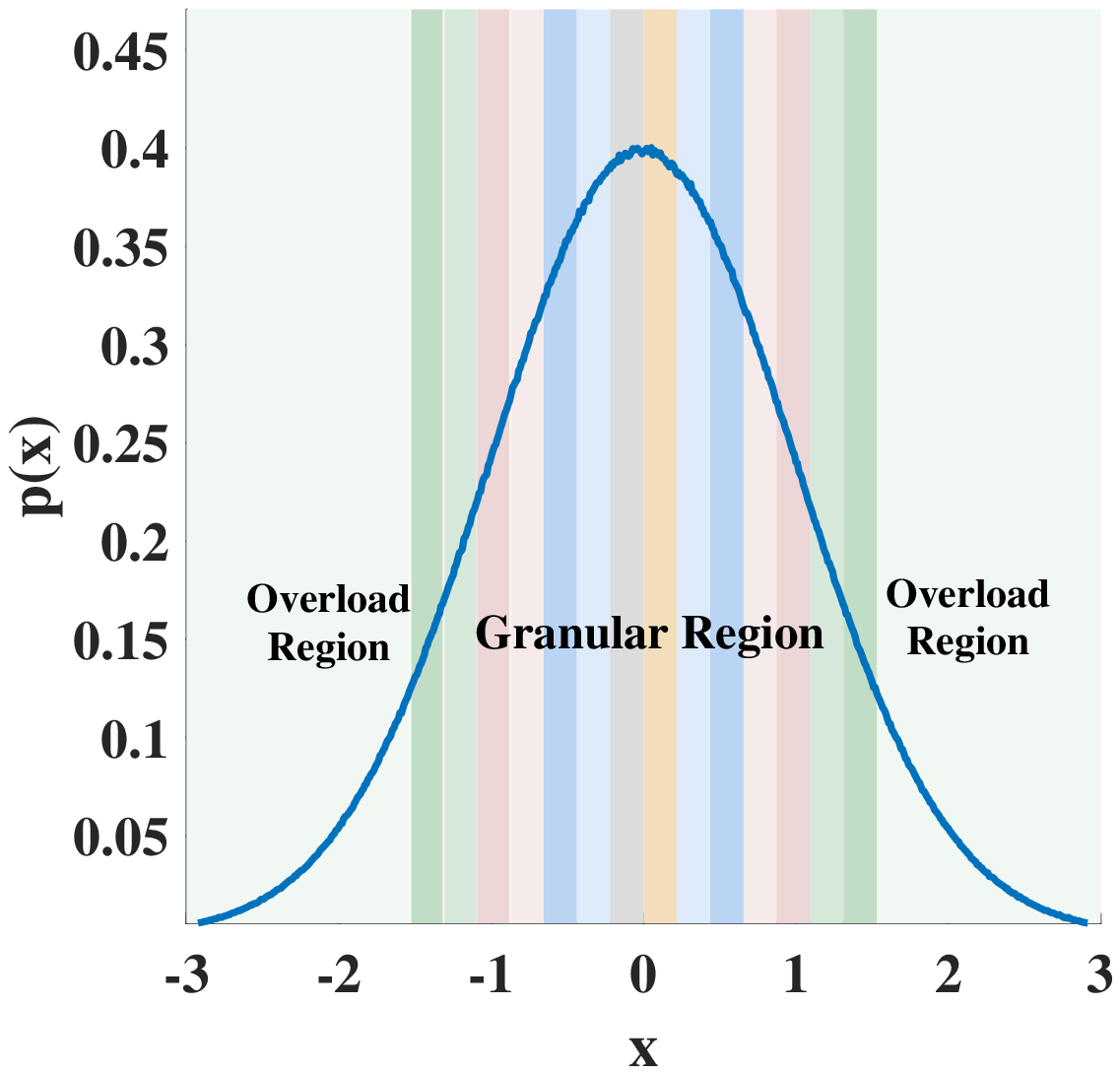}}

  \caption{(a) Third-order nonlinearities characterized by ${P}_{\text{1dB}}$, and histograms of instantaneous input power for passband and baseband signals. (b) Histogram of I and Q baseband components along with ADC quantization bins.} 

  \label{Nonlinearity}
\end{figure}

 We design the quantizer to minimize the mean square error (MSE) assuming that the incoming signal is Gaussian with zero mean and unit variance.  An automatic gain control (AGC) precedes the ADC in order to normalize the average power of the input signal to unity, and hence, ensure that it exploits all the dynamic range.
We employ an overloaded uniform ADC \cite{gersho2012vector}: while the MSE could be improved slightly by designing a non-uniform quantizer for standard Gaussian input, the improvement is slight and has no discernible impact on system-level performance (see Appendix \ref{AppendixA} for a quantitative discussion).  Fig. \ref{Nonlinearity} (b) depicts a 4-bit uniform overloaded quantizer.

\subsection{Linear MMSE Detector}



We show in a following section that the impact of a per-antenna nonlinearity $g(\cdot)$ can be modeled as additional noise, leading
to an equivalent system model of the form
\begin{align} \label{recd2}
\mathbf{y}={\mathbf{H}}\mathbf{x} +\Tilde{ \mathbf{n}},
\end{align} 
where ${\Tilde{\mathbf{n}}} \sim \mathcal{CN}(\mathbf{0},(\sigma_{n}^2+\nu_g^2)\mathbf{I})$, where the value of $\nu_g$ 
is specified in section \ref{analysis}.  Thus, any adaptive implementation of the linear MMSE receiver automatically accounts for the nonlinearities.
The linear MMSE receiver is specified as follows:
\begin{align}
    \hat{\mathbf{x}}=\mathbf{Wy},
\end{align}
where
\begin{align}
    \mathbf{W}=\left(\mathbf{H}^H\mathbf{H}+(\sigma_{n}^2+ \nu_g^2 )\mathbf{I}\right)^{-1}\mathbf{H}^H.
\end{align}

The linear MMSE detector has a rich history with well-known properties \cite{madhow1994mmse, Verdu_book}.  In order to provide a self-contained exposition, we state a few properties that are relevant for our present purpose and sketch their proof in Appendix \ref{AppendixB}.

\section{Bussgang Linearization} \label{BussgangLin}

In order to provide a self-contained exposition, we review Bussgang linearization in the context of our MIMO system.

\subsection{Scalar Bussgang Linearization}

For a zero mean complex-valued random variable $y$ and a nonlinearity $g( \cdot )$, a linear MMSE approximation of $g(y)$ by $ay$
satisfies the orthogonality principle \cite{hajek2015random}:
\begin{equation} 
\mathbb{E}((g(y) - ay) y^*) = 0.
\end{equation}
Standard computations for the linear gain $a$ and the variance of the approximation error $e=g(y)-ay$ yield
\begin{equation} \label{bussgang1}
a = \frac{\mathbb{E}(g(y) y^*)}{\mathbb{E}(|y|^2)},
\end{equation}
\begin{equation} \label{bussgang2}
 \sigma_g^2 = \mathbb{E}(|e|^2) = \mathbb{E}( |g(y) |^2) -  |a|^2 \mathbb{E}(|y|^2).
\end{equation}
Hence, $g(y)$ can be written as
\begin{align}\label{bussgang4}
    g(y) = ay + e,
\end{align}
where $a$ and $\mathbb{E}(|e|^2)=\sigma_g^2$ can be computed analytically or empirically for any distribution of $y$ and nonlinear function $g(\cdot)$. Bussgang evaluated $a$ and $\sigma_g^2$  for different nonlinear functions when the input $y$ is Gaussian random variable \cite{bussgang}. 



\subsection{Vector Bussgang Linearization}

The main part of Bussgang's theorem in \cite{bussgang}, and its extension to the complex domain in \cite{minkoff1985role}, is the
preservation of covariance structure under nonlinearities for jointly Gaussian random variables:\\
If $y$ and $z$ are jointly Gaussian random variables and $g(\cdot)$ is a nonlinear function, then $\mathbb{E}(g(y)z^*)=a\mathbb{E}(y z^*)$, where $a$ is defined in (\ref{bussgang1}).

This result allows us to characterize the linear MMSE fit for a Gaussian random vector in terms of the scalar linear MMSE fits for its components.
It has been customized to MIMO in many recent papers
\cite{bj2018hardware,j2018massive,jacobsson2017throughput,jacobsson2017quantized},
hence we state the relevant result here without proof (see Appendix~A in \cite{jacobsson2017quantized} for a derivation).

\begin{thm} \label{vector_bussgang}
{\bf Vector Bussgang Decomposition}\\
Let $\mathbf{y}$ denotes the jointly Gaussian random vector input to the effective nonlinearity $g( \cdot )$
referred to complex baseband, so that the received signal $\mathbf{z} = g( \mathbf{y})$.
Then the Bussgang decomposition of $\mathbf{z}$ is given by
%
\begin{align}
\mathbf{z} = \mathbf{g}(\mathbf{y})=\mathbf{A}\mathbf{y}+\mathbf{e},
\end{align}
where
\begin{align} \label{bussgang5}
\mathbf{A}=& \mathcal{D}iag([a_1,\hdots,a_N]),\\
a_i =& \frac{\mathbb{E}(g(y_i ) y_i^*)}{\mathbb{E}(|y_i|^2)},
\end{align}
and the variance of element $e_i$ of 
the approximation error vector $\mathbf{e}$ is given by 
\begin{align}\label{bussgang6}
    \sigma_{gi}^2 =& \mathbb{E}( |g(y_i ) |^2) -  |a_i|^2 \mathbb{E}(|y_i|^2).
\end{align}
\end{thm}
The Bussgang theorem on covariance preservation therefore leads to a linear MMSE fit with diagonal structure.  Moreover, the diagonal
elements are equal if the statistics of $\{ y_i \}$ are identical, as in the following straightforward corollary, stated without proof.

\begin{cor}
If the diagonal elements of the covariance of $\mathbf{y}$ are equal, 
i.e., $\mathbb{E}(|y_i|^2)=\mathbb{E}(|y_k|^2),\, \forall \, i , k$, 
then the Bussgang decomposition specializes to
\begin{align}
  \mathbf{z}=  \mathbf{g}(\mathbf{y})=a\mathbf{y}+ \mathbf{e},
\end{align}
where $a$ and $\mathbb{E}(|e_i|^2)=\sigma_g^2$ are the scalar Bussgang parameters of $g(\cdot)$.
\end{cor}

It is worth noting that the self-noise $\mathbf{e}$ may be spatially correlated. However, recent work \cite{bj2018hardware} indicates that 
this correlation becomes negligible when the number of users is large, and we ignore it in our analysis here.

\section{Bussgang Normalization and Intrinsic SNR} \label{BussgangNorm}

In this section, we define a normalization such that the Bussgang parameters for a nonlinearity are independent of input power.
We introduce the concept of {\it intrinsic SNR} to characterize the self-noise in this normalized setting. As we shall see, this is the summary specification that
is provided by system-level design requirements to the hardware designer, based on the analytical framework described in the next section.
Finally, we show, via the simple example of a limiter, how such a summary can be used to determine hardware specifications for a nonlinearity.
 

\subsection*{Normalized Nonlinearity}

\begin{figure}
\centering
\subfloat[Nonlinear model]{\includegraphics[trim=0cm 0 0cm 0,clip,scale=0.4]{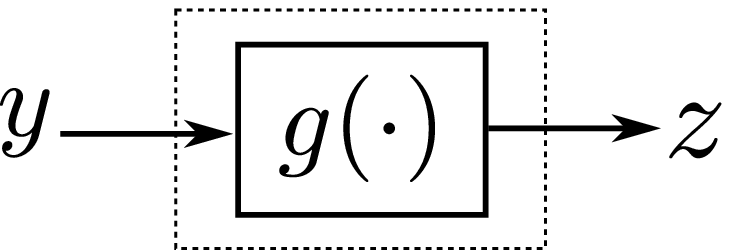}}
\hspace{34 pt}
\subfloat[Linear model]{\includegraphics[trim=0cm 0 0cm 0,clip,scale=0.4]{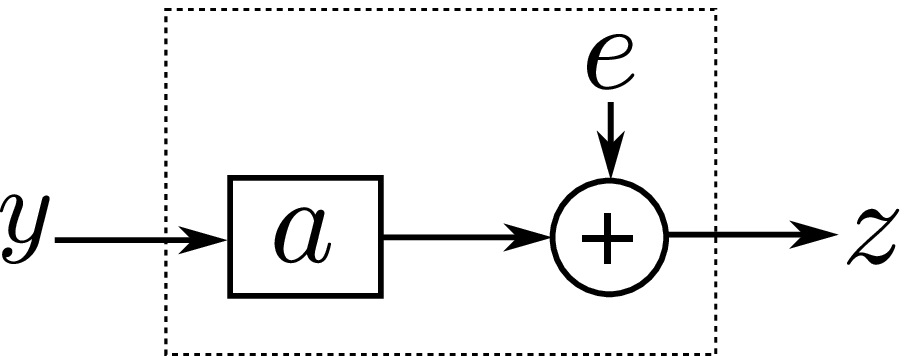}}

\subfloat[Normalized nonlinear model]{\includegraphics[trim=0cm 0 0cm 0,clip,scale=0.4]{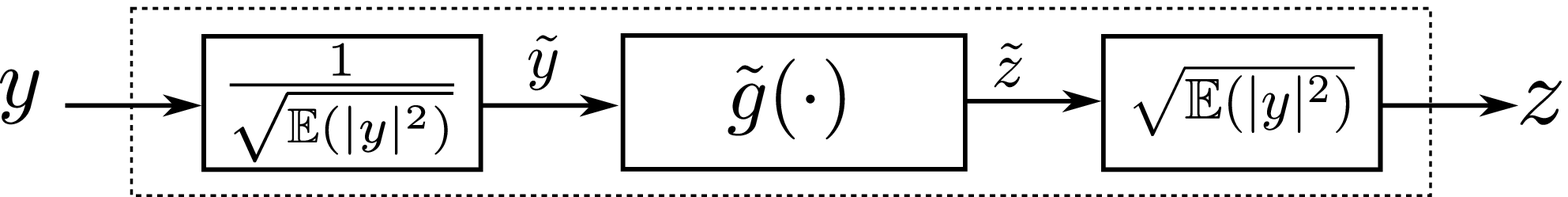}}

\subfloat[Normalized linear model]{\includegraphics[trim=0cm 0 0cm 0,clip,scale=0.4]{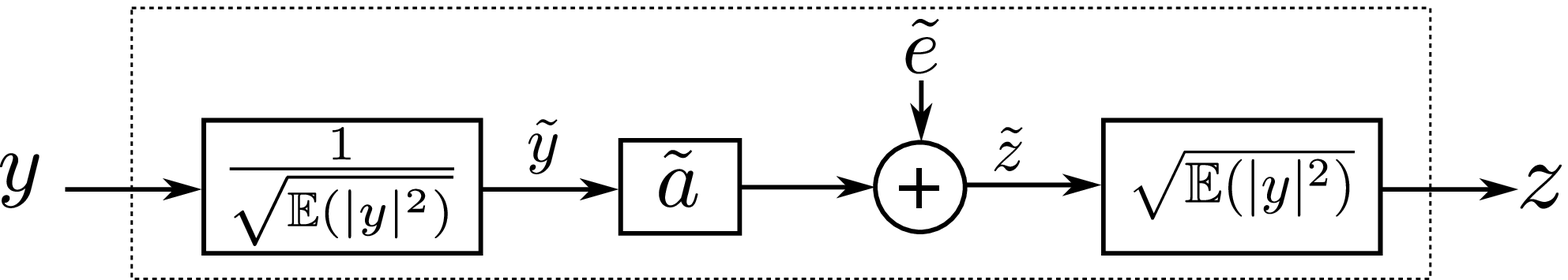}}

\caption{The nonlinear function $g(\cdot)$ in (a) can be decomposed to the linear model in (b) whose parameters depend on the input power. 
We define a normalized version of the nonlinearity in (c), which allows us to provide design specifications independent of input power.
The corresponding normalized linearization is depicted in (d).}
    \label{LMMSEDecompApp}
\end{figure}

As shown in Fig.~\ref{LMMSEDecompApp}~(a) and (b), Bussgang decomposition characterizes a nonlinear function $g(\cdot)$ by parameters $a$ and $\sigma_g^2$. These parameters depend on the input power by definition as shown in Eq.~(\ref{bussgang1}) and (\ref{bussgang2}).


Fig.~\ref{LMMSEDecompApp}~(c) illustrates a normalized version of the nonlinearity in Fig.~\ref{LMMSEDecompApp}~(a): the input power is scaled to one before
the nonlinearity, and the scaling is undone after the nonlinearity. The Bussgang linearization of the normalized nonlinearity,
with parameters $\tilde{a}$ and $\tilde{\sigma}_g^2$, is depicted in Fig.~\ref{LMMSEDecompApp}~(d). 
The parameters $\tilde{a}$ and $\tilde{\sigma}_g^2$ represent the Bussgang decomposition of the {\it normalized} nonlinear function $\tilde{g}(\cdot)$, depicted in Fig.~\ref{LMMSEDecompApp}~(c). The equivalence of the nonlinear models (a) and (c) implies that the corresponding linear models (b) and (d)
must satisfy $\tilde{a}=a$ and $\tilde{\sigma}_g^2=\sigma_g^2/\mathbb{E}(|y|^2)$. 

It is convenient to define hardware specifications for the normalized nonlinearity; in hardware design parlance, the specifications
are "referred to the input power."  We summarize these using the concept of {\it intrinsic SNR,} which plays a key role
in our analytical framework.

\begin{defn}
{\bf Intrinsic SNR}\\
We define the ``{\it intrinsic SNR}" of a nonlinearity $g( \cdot )$ using the Bussgang parameters of its normalized
version $\tilde{g} (\cdot)$ as follows:
\begin{equation} \label{bussgang3}
\gamma_g = \frac{|\tilde{a}|^2}{\tilde{\sigma}_g^2}.
\end{equation}
\end{defn}




\begin{figure}
\centering
\subfloat[Limiter function]{\includegraphics[trim=0cm 0 0cm 0,clip,scale=0.6]{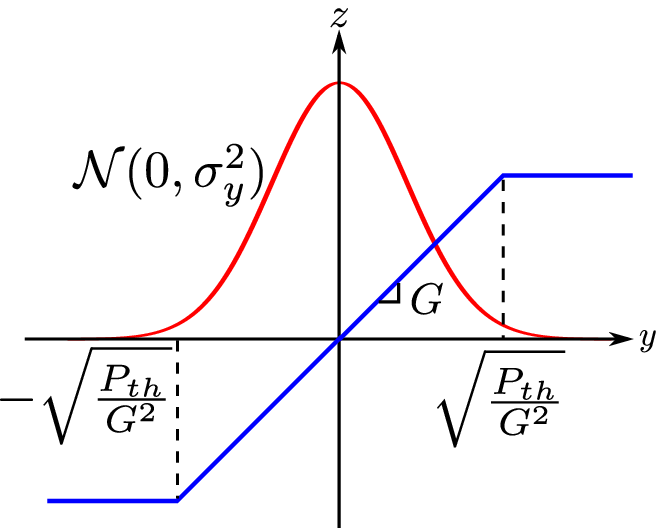}}
\hspace{10 pt}
\subfloat[Normalized limiter function]{\includegraphics[trim=0cm 0 0cm 0,clip,scale=0.6]{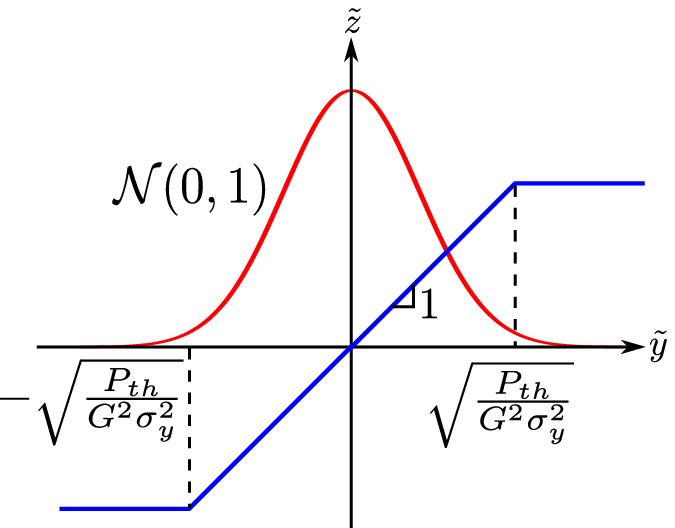}}
\caption{(a) The conventional limiter function. (b) a unity-gain limiter function whose clipping threshold is normalized to the effective input power.}
    \label{LimiterFun}
\end{figure}

\begin{figure}
\centering
\subfloat[Bussgang parameters]{ \includegraphics[trim=25cm 0 21cm 0,clip,scale=0.12]{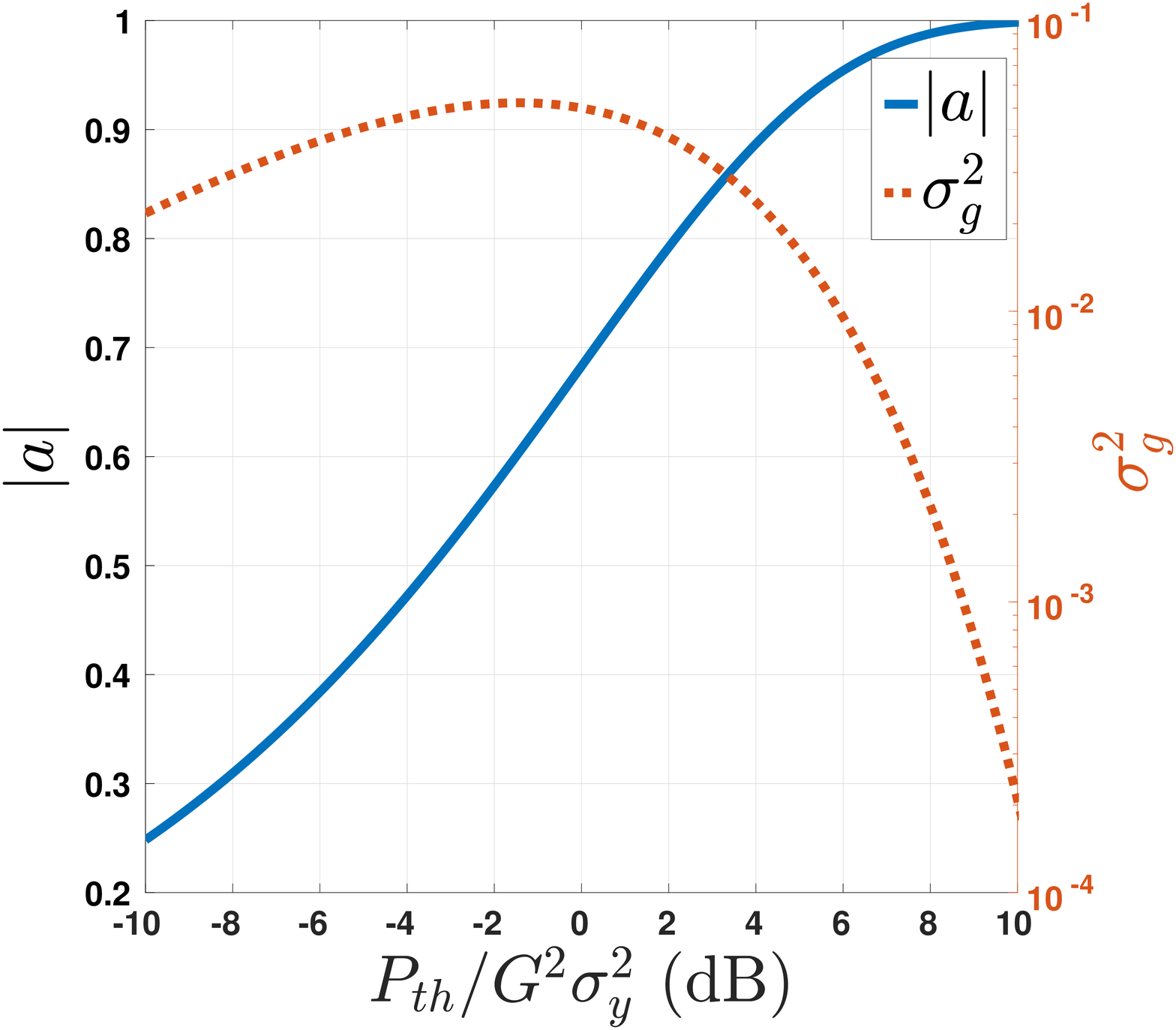}}
\subfloat[Intrinsic SNR]{ \includegraphics[trim=25cm 0 21cm 0,clip,scale=0.12]{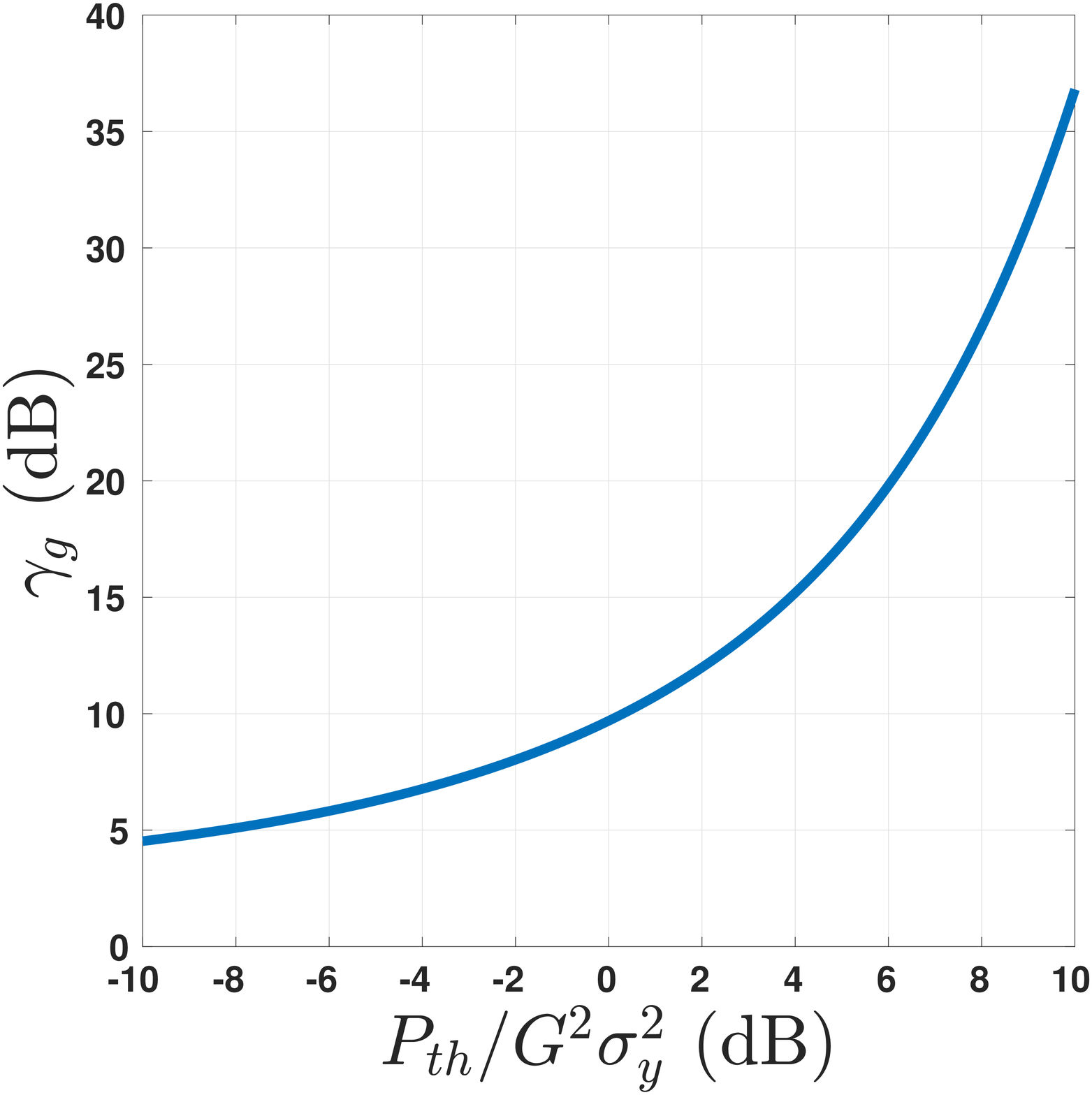}}

\caption{(a) Bussgang  parameters and (b) the intrinsic SNR of the normalized limiter function.}
    \label{IntrinsicSNR}
\end{figure}

As a simple example, consider a memoryless limiter as depicted in Fig.~\ref{LimiterFun}~(a), which is specified by the gain $G$ and the power threshold $P_{th}$ at which the output signal is clipped. The normalized version of this function has unity gain, as shown in Fig.~\ref{LimiterFun}~(b), hence we only need to specify a single
parameter to characterize it: the clipping threshold $\tilde{P}_{th}={{P_{th}}/{G^2 \sigma_y^2}}$ normalized to the input power $\sigma_y^2$.
The Bussgang parameters of the normalized limiter function are shown in Fig.~\ref{IntrinsicSNR}~(a), and the intrinsic SNR is shown in 
Fig.~\ref{IntrinsicSNR}~(b).


Henceforth, nonlinearities and their Bussgang parameters are normalized to the input power, and we drop the ``tilde'' notation
to denote the normalized version. For example, the 1~dB compression point of a passband/baseband nonlinearity is normalized to the input power, and hence
is measured in dB instead of dBm.

\subsection*{Design Approach}

The analytical framework described in the next section leads the following design approach for going from system-level performance metrics
to hardware design specifications:\\
$\bullet$ MIMO performance specifications lead to a requirement for the intrinsic SNR for the per-antenna nonlinearities, 
ignoring the specific nature of the nonlinearities.  For example, suppose that we require an intrinsic SNR of 20 dB at least 95\% of the time.\\
$\bullet$ We map the intrinsic SNR requirement to a specification for the normalized nonlinearity.
Taking the limiter in Fig.~\ref{LimiterFun} as an example, we see from Fig.~\ref{IntrinsicSNR}~(b), the clipping threshold normalized
to the effective input power, ${{P_{th}}/{G^2 \sigma_y^2}}$, must be at least 6 dB in order to attain an intrinsic SNR of 20 dB.\\
$\bullet$ In this step, the absolute value of the gain and clipping threshold is calculated. 
For example, suppose that the system in our running example is at load factor $\beta = 1/4$, i.e., 64 users. Then, according to the link budget presented in Appendix~\ref{Appendix0}, the input power to the receive chain is $-60$~dBm if power control is employed. We therefore obtain that $P_{th}/G^2=-54$~dBm.
The hardware designer now has to choose $G$ and $P_{th}$ in order to achieve this ratio or better.
\section{Analytical Framework} \label{analysis}

Our analytical framework is developed as follows.
\begin{enumerate}
\item 
We derive a matched filter bound for each user in the MIMO system that accounts for the self-noise due to the per-antenna nonlinearities (which scales
with the power summed across users) as well as thermal noise.  To this end, we use Bussgang linearization and the intrinsic SNR
discussed in the previous section.
\item
We derive a lower bound for the output SINR of the LMMSE receiver for any given user.  Defining the {\it efficiency} of the LMMSE receiver for a given user
as the ratio of SINR to SNR, we show that the efficiency of a user in an ideal system without nonlinearities is a {\it lower bound} on that of the actual system.
This, together with the matched filter bound, provides a lower bound on the LMMSE output SINR.
\item
We obtain system-level design prescriptions by specializing the preceding lower bound to an ``edge'' user whose performance is stochastically poorer
than that of any other user.

\end{enumerate}

\subsection{Bussgang Linearized Model}

As described in section \ref{systemmodel}, we denote by $\{ A_k, k=1,\hdots,K  \}$ the amplitudes of the incoming waves for the $K$ users, and by $\sigma_{n}^2$ the variance of the thermal noise at each antenna. We can therefore model the incoming signal at each receive antenna as 
$y_m \sim \mathcal{CN}(0, \sigma_y^2)$, where
\begin{equation} \label{scale}
\sigma_y^2 = \sum_{k=1}^K A_k^2+\sigma_{n}^2 = \sigma_{n}^2 + K A_{rms}^2 ,
\end{equation}
and 
\begin{equation} \label{A_rms}
A_{rms} = \sqrt{\frac{1}{K}  \sum_{k=1}^K A_k^2}
\end{equation}
is the root mean square (rms) amplitude, averaged across users.

As depicted in Fig.~ \ref{LMMSEDecompApp}~(c), using the normalized Bussgang linearization requires scaling the incoming signal to unit variance as follows:
\begin{equation}
\tilde{y}_m = \frac{y_m}{\sigma_y}.
\end{equation}

For a normalized nonlinearity $g( \cdot )$ as defined in the previous section, our per antenna linearized model is given by:
\begin{equation} \label{per_antenna_linearized}
g(\tilde{y}_m) = a \tilde{y}_m + e_m.
\end{equation}

For the received signal (\ref{recd}),
the normalized signal prior to passing through the nonlinearity is given by
\begin{equation} 
\tilde{\mathbf{y}} = \frac{\mathbf{y}}{\sigma_y}.
\end{equation}
Using the Bussgang decomposition, we have
\begin{equation} 
g( \tilde{\mathbf{y}}) = a \tilde{\mathbf{y}} + \mathbf{e} = \frac{a}{\sigma_y} \mathbf{y} + \mathbf{e},
\end{equation}
where $\mathbf{e} \sim \mathcal{CN}(0, \sigma_g^2 \mathbf{I})$.
We can now go back to the original signal scaling to obtain
\begin{equation} \label{linearized_model}
\hat{\mathbf{y}} = \frac{\sigma_y}{a} g( \tilde{\mathbf{y}} ) = \mathbf{y} + \frac{\sigma_y}{a} \mathbf{e} = \mathbf{H} \mathbf{x} + \mathbf{n} + \frac{\sigma_y}{a} \mathbf{e}.
\end{equation}
This is the model (\ref{recd2}), with effective noise 
\begin{equation} \label{effective_noise}
\Tilde{ \mathbf{n}} = \mathbf{n} + \frac{\sigma_y}{a} \mathbf{e} \sim \mathcal{CN}( 0, ( \sigma_n^2 + \nu_g^2 ) \mathbf{I} ),
\end{equation}
where 
\begin{equation} \label{effective_noise2}
\nu_g^2 = \frac{\sigma_y^2}{|a|^2} \sigma_g^2 = \frac{\sigma_y^2}{\gamma_g}.
\end{equation}

\subsection{Matched Filter Bound} 

For the $k^\text{th}$ user, the matched filter bound for the linearized model (\ref{recd2}), with equivalent noise as in (\ref{linearized_model})-(\ref{effective_noise}),
is simply given by 
\begin{equation} \label{MF_bound}
SNR_k (g) = \frac{||{\mathbf{h}}_k||^2}{\sigma_n^2 + \nu_g^2 }.
\end{equation}

Our design framework is built around the dependence of this bound on key system parameters as stated in the following theorem.
We first ignore thermal noise, in order to clearly brings out the role of intrinsic SNR $\gamma_g$ and load factor $\beta$, and then include
its effect. 

\begin{thm} \label{mfb}
{\bf Matched filter bound}\\
(a) {\bf Self-noise only:} Ignoring thermal noise, the matched filter bound for user $k$ is given by
\begin{align}
SNR_k(g) = \gamma_g \frac{A_k^2}{\beta A_{rms}^2}. \label{MFBound}
\end{align}
(b) {\bf Self-noise and thermal noise:} The matched filter bound for user $k$, considering both self-noise and thermal noise, is given by
 \begin{align} \label{SNRg}
SNR_{k}(g,\sigma_{n}^2) =& \frac{1}{ \frac{1}{ SNR_{k}(g)}+\frac{1+\gamma_g}{\gamma_g}\frac{1}{ SNR_{k}}},
\end{align}
where $SNR_{k} = {NA_{k}^2}/{\sigma_{n}^2}$ is the SNR for user $k$ accounting for thermal noise alone.
\end{thm}

\begin{proof}
The proof involves algebraic manipulations based on the linearized model (\ref{linearized_model})-(\ref{effective_noise}).  \\
(a) Using (\ref{LoS_channel}), the numerator in (\ref{MF_bound}) is given by
\begin{equation} \label{MF1}
||{\mathbf{h}}_k||^2 = N A_k^2.
\end{equation}
Using (\ref{scale}) and (\ref{effective_noise2}), and setting $\sigma_n^2 = 0$, the denominator in (\ref{MF_bound}) is given by
\begin{equation} \label{MF2}
\nu_g^2 = \frac{K A_{rms}^2}{\gamma_g}.
\end{equation}
Plugging (\ref{MF1}) and (\ref{MF2}) into (\ref{MF_bound}), we obtain
\begin{equation} \label{MF3}
SNR_k (g) = \frac{N A_k^2 \gamma_g}{K A_{rms}^2} = \frac{\gamma_g A_k^2 }{\beta A_{rms}^2},
\end{equation}
which is the desired result (\ref{MFBound}).\\
(b) From (\ref{MF_bound}) and (\ref{MF1}), we have 
\begin{equation} \label{MF4}
\frac{1}{SNR_k (g,\sigma_{n}^2)} = \frac{\sigma_n^2}{N A_k^2} + \frac{\nu_g^2}{NA_k^2}.
\end{equation}
For non-zero thermal noise, we have, using (\ref{scale}) and (\ref{effective_noise2}), that
\begin{equation} \label{MF5}
\nu_g^2 = \frac{K A_{rms}^2 + \sigma_n^2}{\gamma_g}.
\end{equation}
Plugging into (\ref{MF4}), we obtain upon simplification the desired result (\ref{SNRg}).

\end{proof}

Note that, if $\gamma_g \gg 1$, then the formula (\ref{SNRg}) reduces to 
 \begin{align}
SNR_{k}(g,\sigma_{n}^2) =& \frac{1}{ \frac{1}{ SNR_{k}(g)}+\frac{1}{ SNR_{k}}}.\label{SNRgSigma}
\end{align}

In order to provide system-level performance guarantees, we focus on supporting users at the cell edge.
We therefore now set $A_k$ to the worst-case amplitude $A_{edge}$ (at 100 m range for our running example), while computing 
$A_{rms}$ by a statistical average $\sqrt{\mathbb{E}[A^2]}$ given the users distribution, assuming a large enough number of users.  
Let us term the ratio of the power of the edge user to the rms power as the {\it power control factor,} since it depends on the power control
scheme used. The power control factor $\alpha_p$ is given by
\begin{equation} \label{power_control_factor}
\alpha_p = \frac{A_{edge}^2}{ A_{rms}^2}.
\end{equation}
Specializing (\ref{MFBound}) to the edge user, we now obtain that
\begin{align}
SNR_{edge}(g) = \gamma_g ~ \frac{1}{\beta} ~ \alpha_p.
\end{align}

\noindent
{\it Power control factor with no power control:} For users who are uniformly distributed over the area bounded by $[R_{min}, R_{max}]$ and a given angular range, we obtain upon straightforward computation that, for a system without power control,
\begin{align} \label{no_power_control}
\nonumber \alpha_p =& \frac{\frac{1}{R^2_{max}}}{\frac{1}{R^2_{max}-R^2_{min}}\int_{R^2_{min}}^{R^2_{max}}\frac{1}{r}dr},\\
 =& \frac{1 - \frac{R_{min}^2}{R_{max}^2}}{2 \log \frac{R_{max}}{R_{min}}}.
\end{align}
which evaluates to -7.8~dB for $R_{max} = 100$~m, $R_{min} = 5$~m.

\subsection{Lower Bound on LMMSE Output SINR}

We now provide a lower bound on the output SINR of any user via the ideal system.  
\begin{thm}
{\bf LMMSE Lower Bound}\\
In the presence of nonlinearity, a lower bound on the output SINR of a linear MMSE for any user is given as
\begin{align}
  SINR \geq& ~ SNR(g,\sigma_{n}^2) ~ \eta_{ideal}, \label{LMMSEBoundeq}
\end{align}
where
\begin{align}
\eta_{ideal}=& \frac{SINR(Ideal)}{SNR(Ideal)}. \label{etaIdeal}
\end{align}
is the efficiency in an ideal system with the same user configuration and amplitudes, but without nonlinearity.
\end{thm}

\begin{proof}
Since the effective noise is higher in the system (\ref{recd2}) than in (\ref{recd}), we have by Lemma C.1 in Appendix \ref{AppendixB} that
\begin{align}
   \frac{SINR_{}}{SNR_{}(g,\sigma^2_n)}\geq \frac{SINR_{}(Ideal)}{SNR_{}(Ideal)},
\end{align}
where ${SINR_{}(Ideal)}$ is the target linear MMSE output SINR for a user in an ideal system (without nonlinearities). 
\end{proof}

\begin{figure}
\centering
\subfloat[BER in ideal system]{ \includegraphics[trim=7.5cm 0 7.8cm 0,clip,scale=0.35]{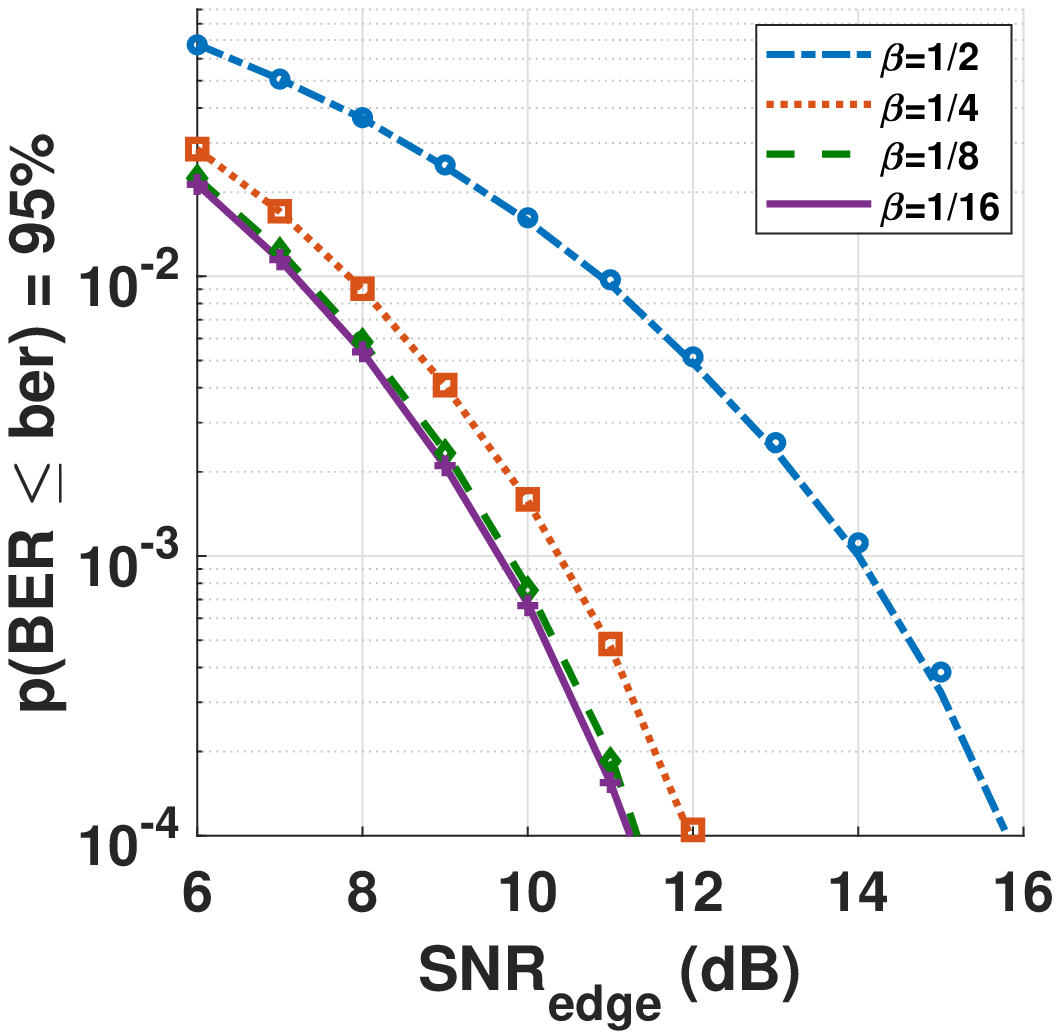}}
\subfloat[$SNR_{edge}$ required in ideal system]{ \includegraphics[trim=7.5cm 0 7.8cm 0,clip,scale=0.35]{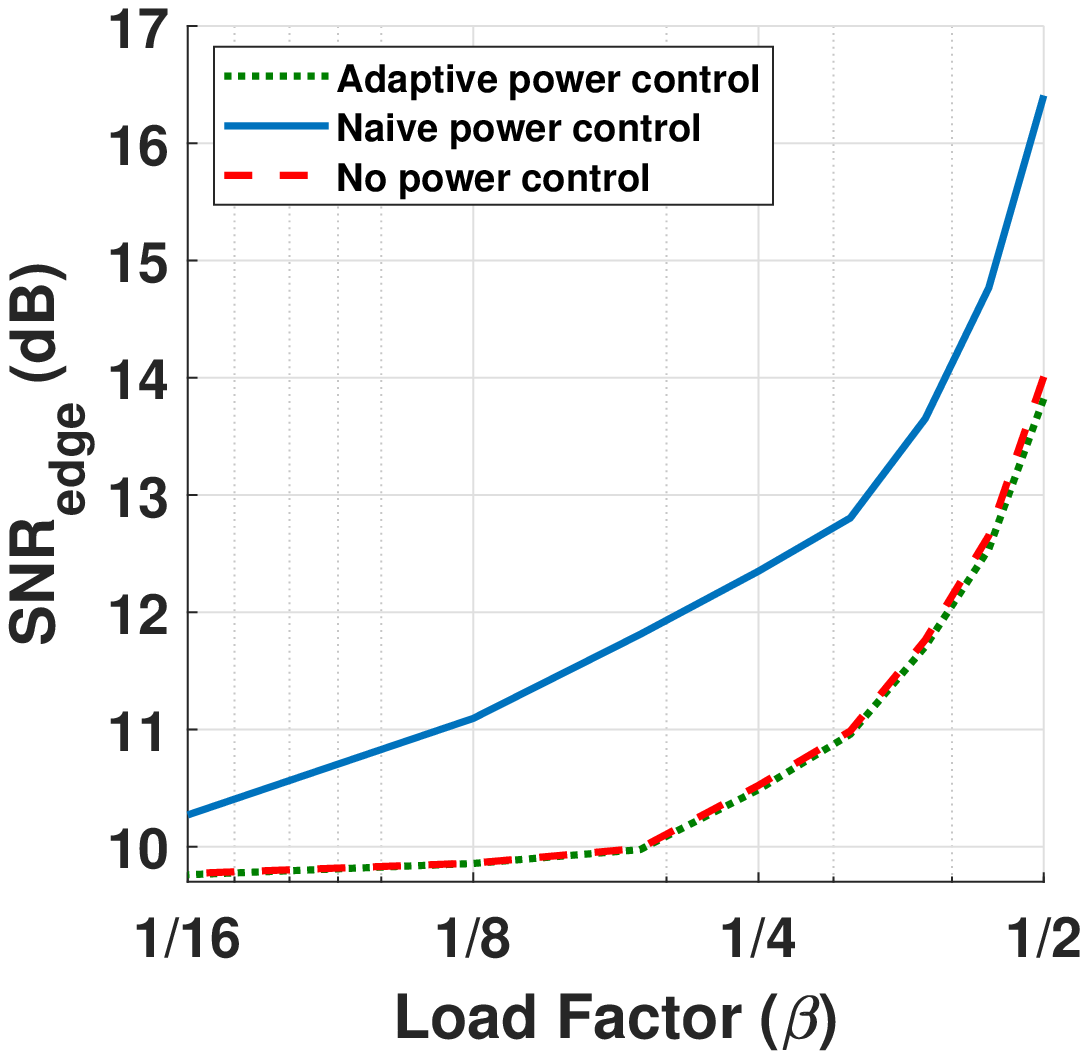}}
\caption{(a) BER for 5\% outage in an ideal system (no nonlinearities) for different load factors. (b) SNR for an edge user (100 m from base station) to guarantee that 95\% of the mobiles have raw BER of $10^{-3}$ for different load factors. }
    \label{SINRGap}
\end{figure}

We evaluate $\eta_{ideal}$ through simulations of the ideal system for the edge user, as shown in Fig. \ref{SINRGap}, where $SNR_{edge}=\frac{NA^2_{edge}}{\sigma_{n}^2}$, and $A_{edge}$ is the received amplitude of the user at 100 m.
The target output SINR of the linear MMSE, i.e., $SINR_{edge}(ideal)$ is 9.7 dB. This number corresponds to the $SNR_{edge}$ in a single user case. Hence, in a single user case $\eta_{ideal}=1$. As the load factor increases, there is noise enhancement due to interference suppression: 
$\eta_{ideal} = 9.7 - SNR_{edge}|_{db}$ can be inferred from Fig. \ref{SINRGap} (b). 

\subsection{From System-Level Performance to Intrinsic SNR}

The chosen quality of service measure maps to an SINR requirement at the LMMSE output.  We compute this for the ideal system.
For example, simulating the ideal system, a target BER of $10^{-3}$ with 95\% availability is obtained
for $SINR_{edge}(ideal) = 9.7$ dB.  Since the SNR for an edge user is 14 dB, we see from Fig. \ref{SINRGap}~(b) that the efficiency for the ideal system is given by $9.7-14=-4.3$~dB 
for no power control and $\beta=1/2$.  This is an upper bound on the efficiency of the actual system.

We can now compute the minimum $SNR_{edge}(g,\sigma_n^2)$ from Eq. (\ref{LMMSEBoundeq}) 
to achieve the required SINR in the presence of nonlinearities. Finally, we can infer the intrinsic SNR $\gamma_g$ required from Eq. (\ref{SNRg}) and Eq. (\ref{SNRgSigma}).  This is now mapped to detailed hardware specifications, as illustrated by examples in the next section.

\section{Design Examples and Performance Evaluation}

The system parameters are as described in Section  \ref{systemmodel}. 
We illustrate our design for a target uncoded BER of $10^{-3}$, which is low enough for reliable performance using a high-rate channel 
code with relatively low decoding complexity.  For QPSK, the corresponding required SNR over a SISO AWGN link is 9.7 dB.  This becomes our target
SINR at the output of the LMMSE receiver for an edge user.  This setting is simply for illustration: our analytical framework applies for any QoS
measure that can be approximated in terms of SINR (e.g., outage capacity or spectral efficiency using Shannon's formula).

In the following, we first describe the user distribution and power control schemes deployed in the cell.
Then, we apply the analytical design framework to define the specification on the receive chain: the passband/baseband nonlinearity and the ADC resolution.  We then evaluate the efficacy of the framework in attaining the desired system-level performance by simulations for selected scenarios.
Finally, we provide design guidelines on the receive chain requirements in a more comprehensive set of scenarios.
\subsection{User Distribution}
\begin{figure}
\centering
\subfloat[Example distribution of mobiles]{ \includegraphics[trim=0cm 0 0cm 0,clip,scale=0.35]{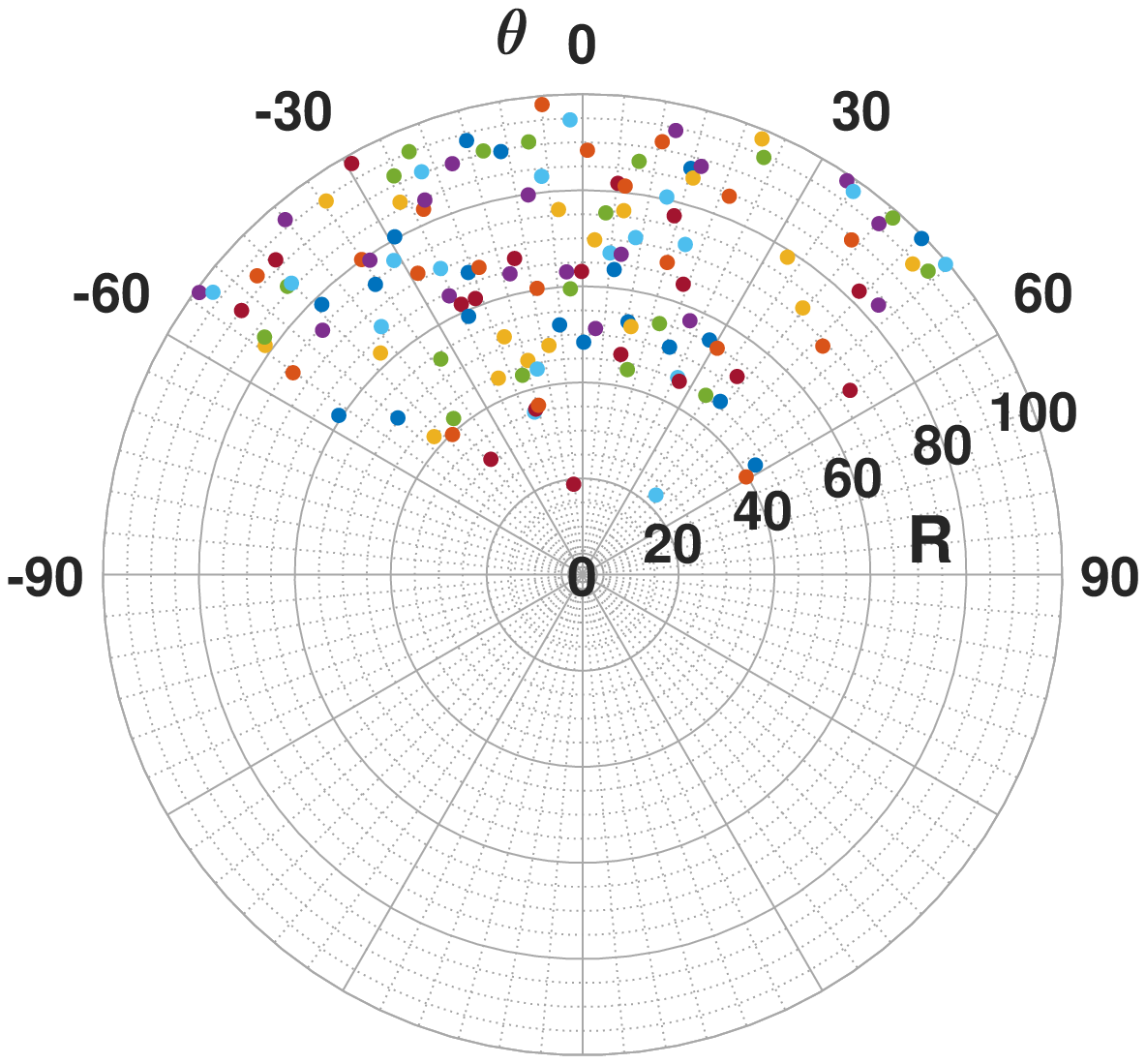}}
\subfloat[Normalized spatial cross-correlation]{ \includegraphics[trim=7cm 0 7cm 0,clip,scale=0.35]{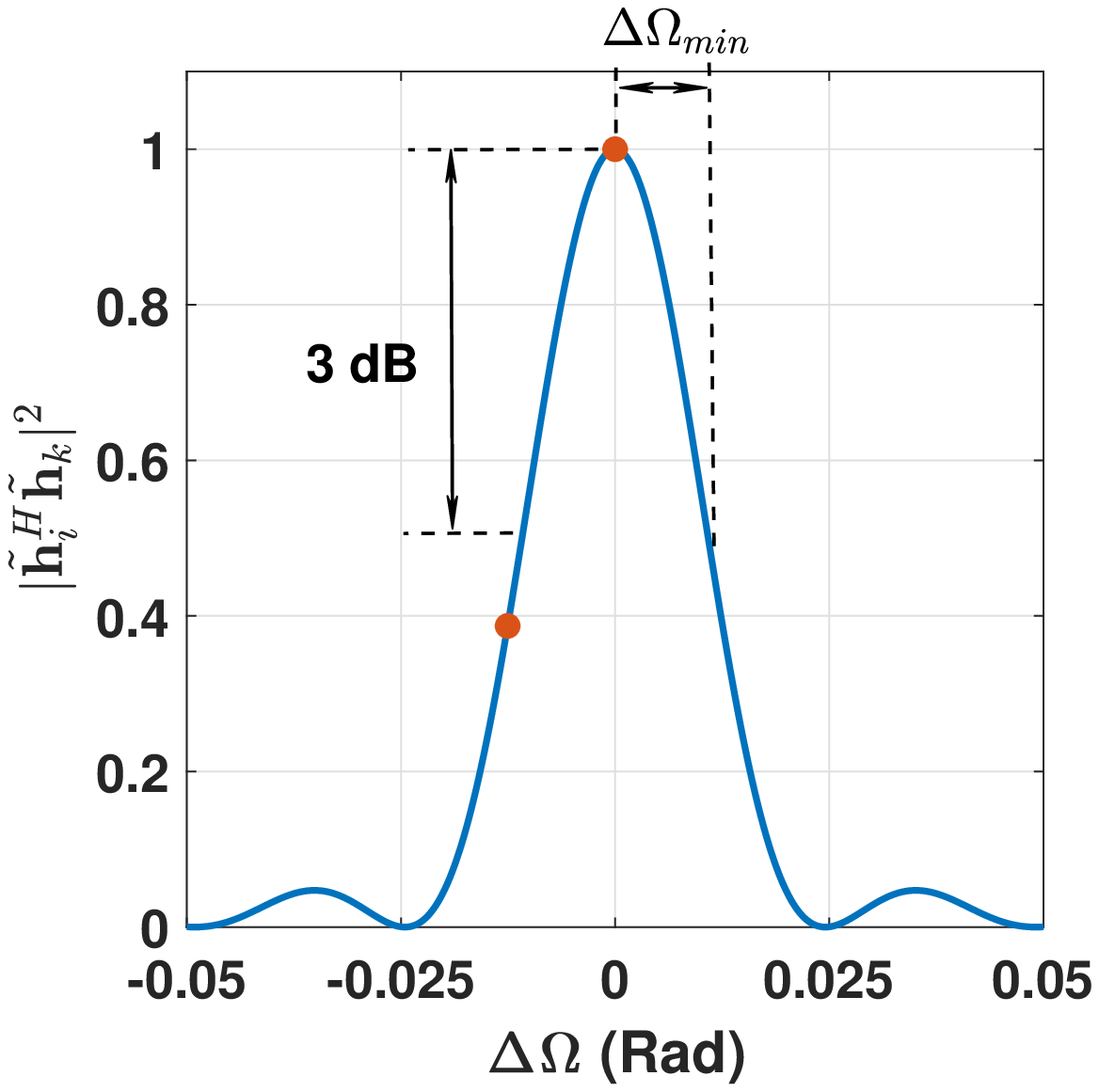}}
  \caption{(a) An instantiation of 128 mobiles on a polar chart. 
(b) Normalized correlation between two users with spatial frequency difference of $\Delta \Omega$.
Note that the closest users, depicted by red points, are separated by larger or equal to half the 3~dB beamwidth.}
  \label{UserDistXcorr}
\end{figure}

The mobiles are uniformly distributed inside a region bordered by a minimum and a maximum distance away from the base station,  $R_{\text{min}}$ and $R_{\text{max}}$, respectively.  Since $\frac{d \Omega}{ d \theta } \sim \cos \theta$, the spatial frequency is less responsive to changes in DoA for $\theta$ near $\pm \frac{\pi}{2}$, which makes
it more difficult to separate mobiles towards the edge of the angular field of view.  We therefore confine the field of view for the antenna array to  $-\pi/3\leq\theta \leq \pi/3$.
While the mobiles are placed randomly in our simulations, 
we enforce a minimum separation in spatial frequency between any two mobiles in order not to incur excessive interference, choosing it as half the 3~dB beamwidth: $\Delta \Omega_{\text{min}} = \frac{2.783}{N}$  \cite{Balanis} (mobiles closer in spatial frequency could be served in different time slots, for example). An example distribution of mobiles is depicted in Fig.~\ref{UserDistXcorr}.

\subsection{Power Control Schemes}

Our analysis in Section \ref{analysis} first considers a system with no power control, in which each transmitter transmits at equal power.  We then consider two power control schemes:
a naive scheme in which transmitters adjust their powers to be roughly equal at the receiver, to within a tolerance, and an adaptive power control scheme aimed at meeting an SINR target for each mobile at the receiver. Power control is a very well-studied area, hence our goal is to provide quick insight on its implications for our system, rather than performing a comprehensive evaluation.  

\subsubsection{Naive power control}
In this scheme, the base station asks all the users to decrease their power to make their received power at the base station equal the received power of the farthest mobile, i.e., at $R_{max}$. A disadvantage of this scheme, illustrated by
our performance results in subsequent subsections, is that nearby users are no longer able to use their larger signal strength to overcome the impact of interference from other users who are nearby (in terms of spatial frequency separation). 
The power factor $\alpha_p$ of the naive power control scheme is equal to $0$~dB because all the users have the same received signal strength.

\subsubsection{Adaptive power control} \label{sec:adaptive_power_control}

In order to avoid the pitfalls of naive power control, we consider an adaptive power control scheme (Algorithm \ref{Algorithm}) aimed at meeting an SINR target $\text{SINR}_{th}$ at the output of the linear MMSE receiver \cite{UlukusYates}. Starting from no power control and all users transmitting
at maximum power, the algorithm seeks to enforce a threshold SINR,
termed $\text{SINR}_{th}$, iteratively as follows: every mobile with SINR greater than $\text{SINR}_{th}$ reduces its power by $\text{SINR} - \text{SINR}_{th}$. The process, specified in Algorithm \ref{Algorithm}, is repeated up to a maximum number of iterations $\text{nIter}$, or until a convergence criterion is met, whichever comes earlier.
The power factor $\alpha_p$ of the adaptive power control scheme can be computed using simulation, and equals about $-2$~dB.

\begin{algorithm}[h]
\KwIn{$\mathbf{H}$,$P^{(0)}_k$ $\forall$  $k\in [1,K]$}
\Parameter{$\text{SINR}_{th}$, nIter}
\KwOut{$P^{(\text{nIter})}_k$}

\nl \For{$i\gets1$ \KwTo $\text{nIter}$ }{
\nl ${SINR}_k$ $\gets$  calculate the LMMSE output SINR\;
\nl  $\Delta{SINR}_k$ $\gets$ $\max({SINR}_k-\text{SINR}_{th},0)$\;
\nl ${P}^{(i)}_k$ $\gets$ ${P}^{(i-1)}_k$-$\Delta{SINR}_k$\;
    }
    \caption{{ Adaptive power control} \label{Algorithm}}
\end{algorithm}

\subsection{Applying the Design Framework}

For illustration, we consider four scenarios: (a) no power control, $\beta = \frac{1}{2}$, (b)  no power control, $\beta = \frac{1}{16}$,
(c) adaptive power control, $\beta = \frac{1}{2}$, (d)  adaptive power control, $\beta = \frac{1}{16}$.

\begin{figure}
\centering
\subfloat[Contours of $SINR_{edge}$]{ \includegraphics[trim=7cm 0 7.8cm 0,clip,scale=0.35]{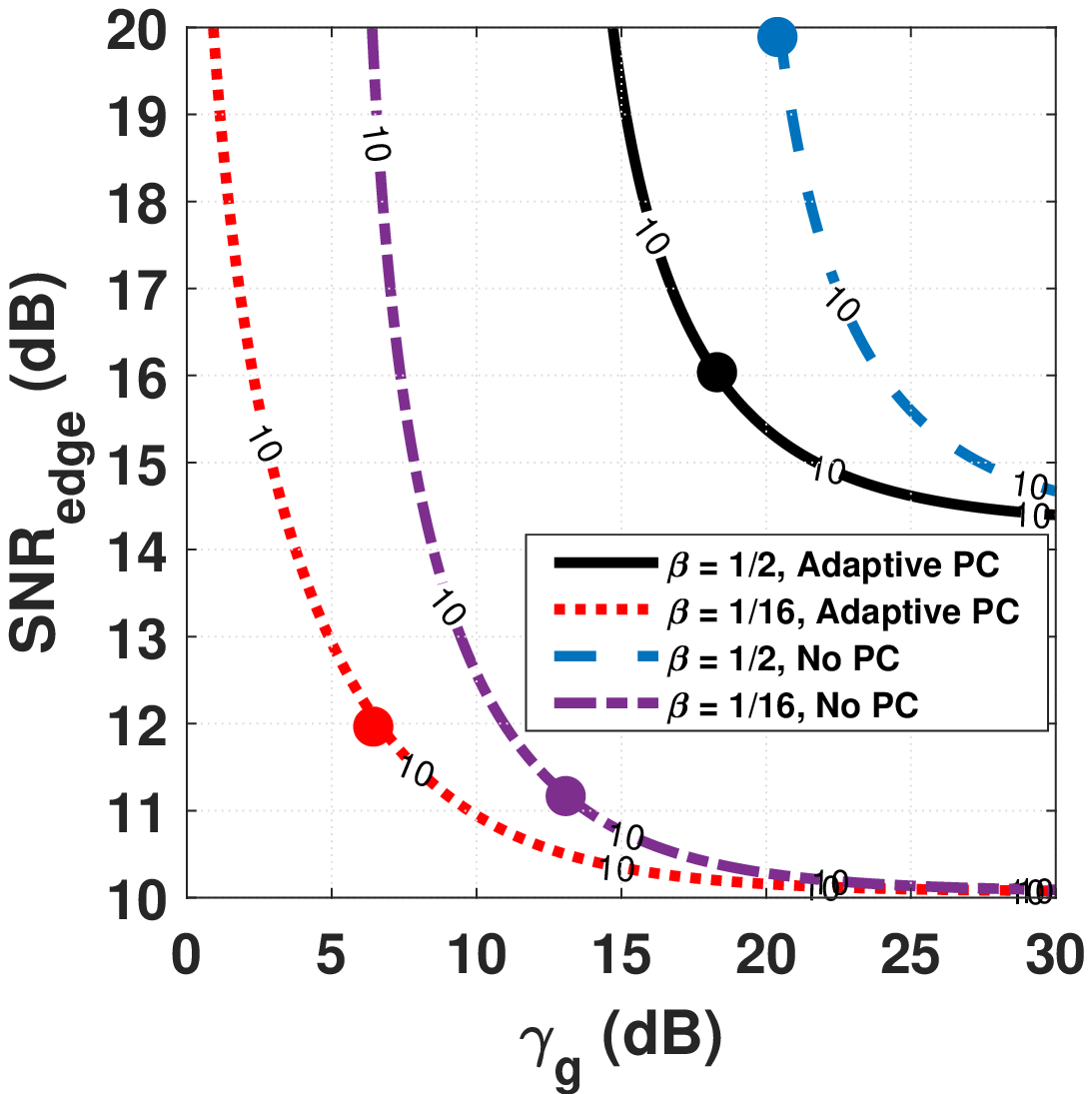}}
\subfloat[Contours of intrinsic SNR $\gamma_g$]{ \includegraphics[trim=7.5cm 0 7.8cm 0,clip,scale=0.35]{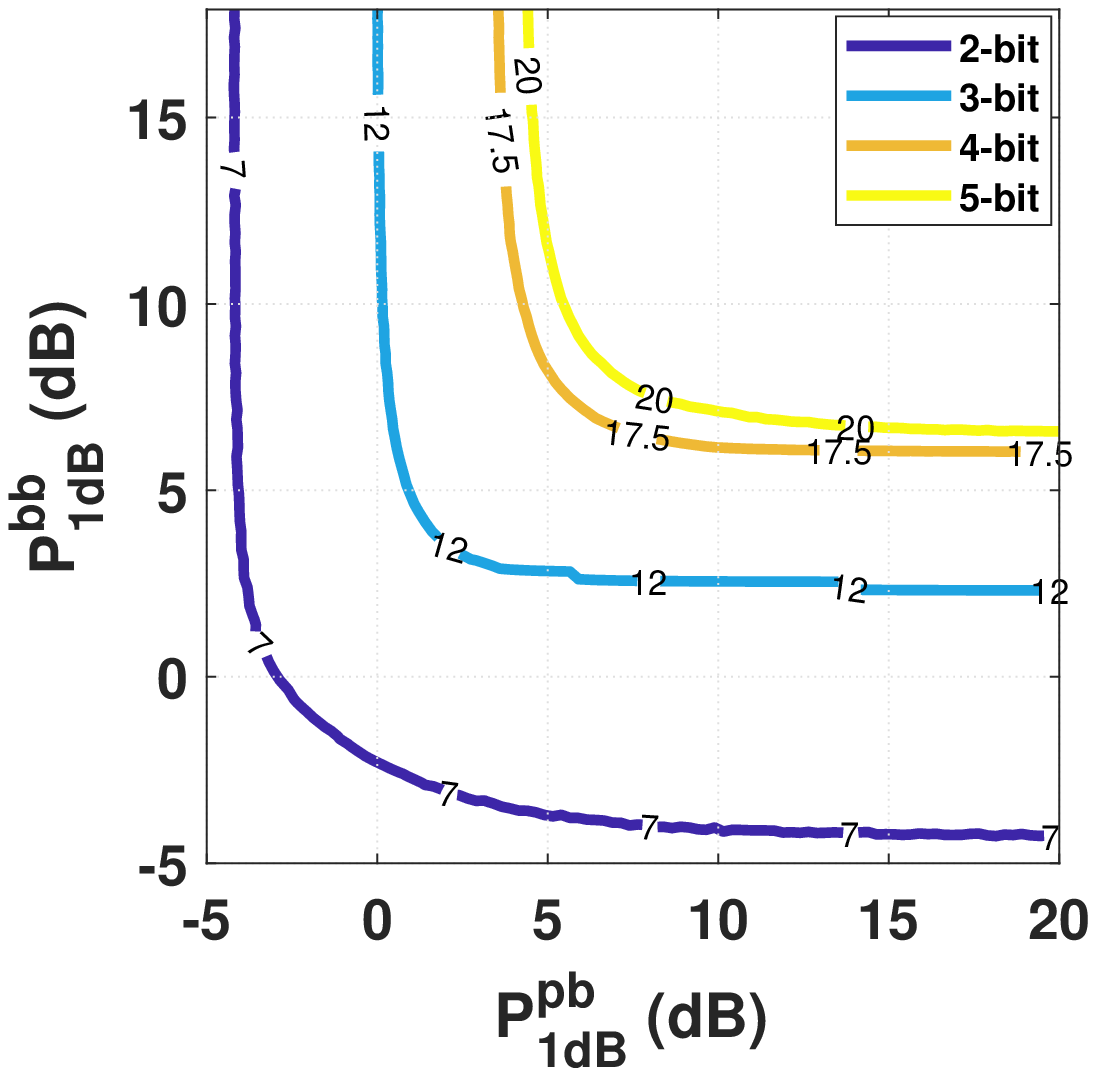}}

\caption{(a) Lower bound on the linear MMSE output SINR as a function in the intrinsic SNR $\gamma_g$ and the SNR required for the edge user $SNR_{edge}$ for different scenarios. The contours depicted are for constant $SINR_{edge}=10$~dB. The solid circles in Fig. (a) show the operating points we choose to work at. (b) Intrinsic SNR of a receive chain comprising passband and baseband nonlinearities and ADC.}
    \label{SNRIntirinsicvsEdge}
\end{figure}

The design steps are as follows:
\begin{enumerate}
\item System-level  design:  We require $SINR_{edge}(ideal)\approx10$~dB for our target QoS. Using simulations for the ideal system, we compute the LMMSE efficiency $\eta_{ideal}$ as shown in Fig. \ref{SINRGap}.
For our four scenarios, the LMMSE efficiency $\eta_{ideal}$ is found to be (a) 4.5~dB, (b) 0~dB, (c) 4.5~dB, and (d) 0~dB.

After that, we determine the SNR of the edge mobile and the intrinsic SNR jointly to achieve the LMMSE lower bound. Specifically, the contours in Fig.~\ref{SNRIntirinsicvsEdge}~(a) illustrates the following equation for each scenario:
\begin{align*}
SNR(g,\sigma_n^2) = \frac{SINR_{edge}}{\eta_{ideal}},\\
\frac{1}{\frac{\beta}{\gamma_g \alpha_p}+\frac{1+\gamma_g}{\gamma_g} \frac{1}{SNR_{edge}}} = \frac{10}{\eta_{ideal}}.
\end{align*}
We pick the following combinations of ($SNR_{edge}$,$\gamma_g$): (a) (20,20) dB, (b) (11,12) dB, (c) (16,17.5) dB, and (d) (12,7) dB.

\item Hardware-level  design: This step determines the specifications of the passband/basband nonlinearity and the ADC  to achieve the required intrinsic SNR. Fig. \ref{SNRIntirinsicvsEdge}~(b) shows the trade-off between the number of ADC bits and the 1~dB compression point of the baseband nonlinearity $P^{\text{bb}}_{1\text{dB}}$ and the passband nonlinearity $P^{\text{pb}}_{1\text{dB}}$.
The 1~dB compression point computed are normalized to the input power. The absolute compression points in dBm are computed by determining the
average received input power at each base station antenna.
\end{enumerate}

Here we have taken the link budget, or attainable $SNR_{edge}$, as our constraint, and have designed the nonlinearity specifications accordingly,
The same framework, of course, also allows us to determine the link budget required for a given set of nonlinearities.

\begin{figure}
\centering
\subfloat[BER with load factor of 1/2]{ \includegraphics[trim=7.5cm 0 7.8cm 0,clip,scale=0.35]{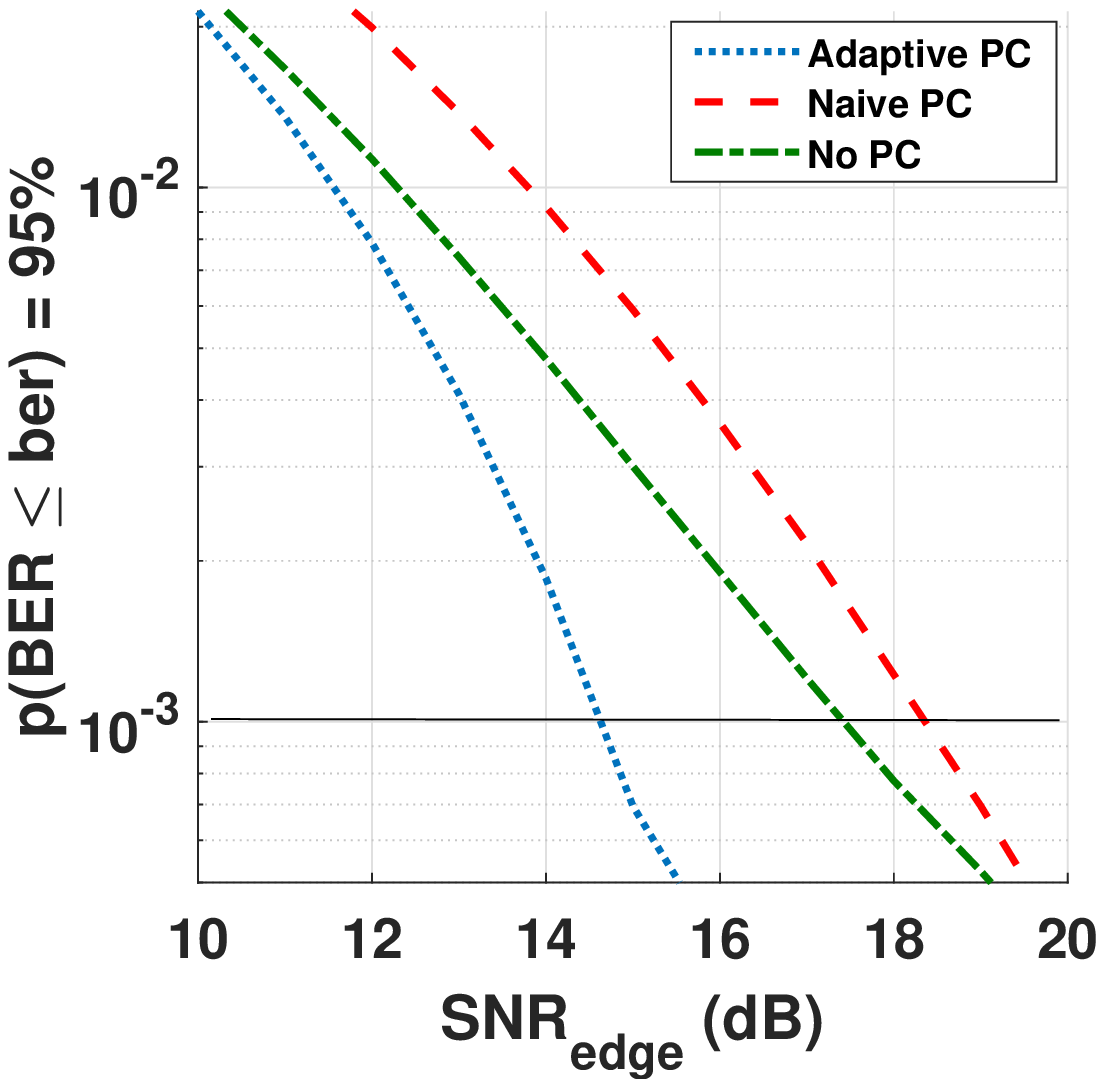}}
\subfloat[BER with load factor of 1/16]{ \includegraphics[trim=7.5cm 0 7.8cm 0,clip,scale=0.35]{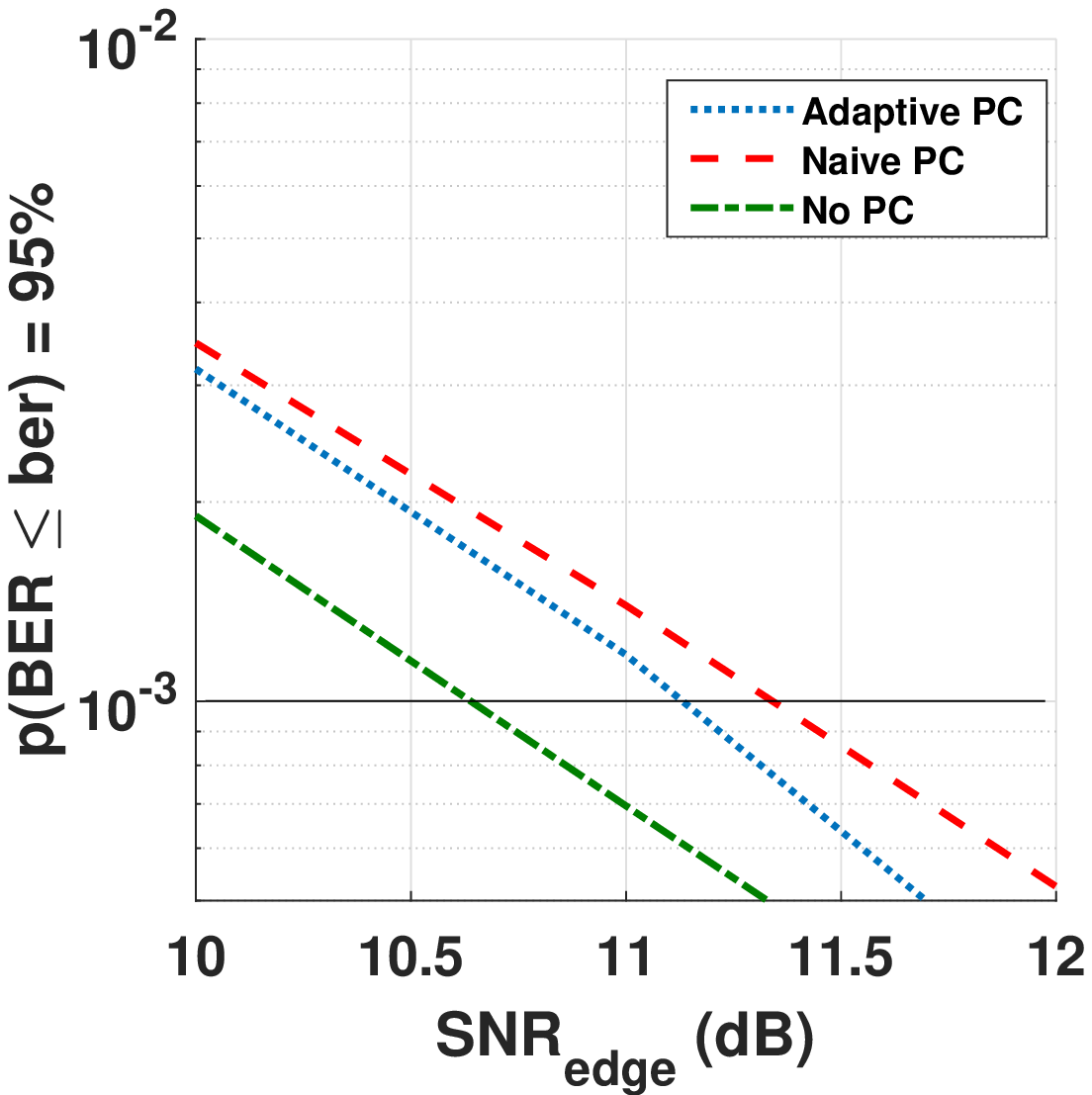}}

\caption{(a) and (b) show the BER attained by 95\% of the users for load factor of 1/2 and 1/16, respectively. The $SNR_{edge}$ is the SNR required by the user at 100 m away from the base station. The receive chain specifications for each curve are demonstrated in table \ref{ResultsSummary}.}
    \label{BER}
\end{figure}

\begin{table}[]
\centering
\caption{This table presents the analytical predictions and simulation results for the SNR budget needed to meet the desired
performance criterion ($10^{-3}$ BER at 95\% availability) for different scenarios. The intrinsic SNR $\gamma_g$ corresponds
to the cascade of the passband and baseband nonlinearities, specified by their 1~dB compression points  
($P^{\text{pb}}_{1\text{dB}}$ and $P^{\text{bb}}_{1\text{dB}}$, respectively), together with $b$-bit ADCs for I and Q. PC and $\beta$ denote the power control scheme used, 
and the load factor, respectively.}
\label{ResultsSummary}
\begin{tabular}{|c|c|c|c|c|c|c|c|}
\hline
$\beta$ & PC       & b & $P^{\text{bb}}_{\text{1dB}}$ & $P^{\text{pb}}_{\text{1dB}}$ & $\gamma_g$ & $SNR_{edge}$  & $SNR_{edge}$ \\
	& 	      & 	&										(dB)		& 						(dB)						& 				(dB)	& (upper bound) & (sim.) \\ \hline
1/2     & none     & 5 & 8.4                          & 6.7                          & 20              & 20                                             & 17.5                       \\ \hline
1/2     & naive    & 4 & 8.4                          & 4.9                          & 17.5            & 18.7                                           & 18.4                       \\ \hline
1/2     & adaptive & 4 & 8.4                          & 4.9                          & 17.5            & 16                                           & 14.7                       \\ \hline
1/4     & none     & 4 & 8.2                          & 2.4                          & 15              & 14                                             & 12.8                       \\ \hline
1/4     & naive    & 3 & 3.7                          & 0.7                          & 10.5            & 15                                             & 14.8                       \\ \hline
1/4     & adaptive & 3 & 3.4                          & 1.9                          & 11.5            & 12.5                                           & 11.9                       \\ \hline
1/8     & none     & 3 & 4.2                          & 1.4                          & 12              & 13                                             & 12.2                       \\ \hline
1/8     & naive    & 3 & 2.2                          & -1.1                         & 8.7             & 12.7                                           & 12.7                       \\ \hline
1/8     & adaptive & 3 & 3.2                          & 1.9                          & 11              & 10.9                                           & 10.8                       \\ \hline
1/16    & none     & 3 & 4.2                          & 1.4                          & 12              & 11.2                                           & 10.8                       \\ \hline
1/16    & naive    & 2 & 1.4                          & -1.1                         & 7.6             & 11.5                                           & 11.5                       \\ \hline
1/16    & adaptive & 2 & -1.1                          & -1.9                          & 7             & 11.8                                             & 11.2                       \\ \hline
\end{tabular}
\end{table}

\subsection{Simulation-based Verification}

Here, we verify the designs produced by our analytical framework by numerical simulations. In Fig. \ref{BER}, we plot the BER that 95\% of the users attain for the cases we mention in the previous subsection. As shown, all the curves reach the $10^{-3}$ at slightly smaller $SNR_{edge}$ than
predicted by our analytical framework, which shows that our approach is both conservative and accurate.
 
Table \ref{ResultsSummary} summarizes our design prescriptions for different scenarios. As shown in the table, we examine the combination of four load factors with no power control and two power control strategies. We demonstrate the specification of the receive chain along with the resultant intrinsic SNR $\gamma_g$. Then we compute an upper bound for the SNR needed for the edge user to achieve the performance metric. Finally, using simulations, we show the accuracy of the derived upper bound. It is worth noticing that power control relaxes the requirements on the receive chain significantly, as predicted by our analytical framework.

\section{Conclusion}

The analytical framework provided in this paper is a conservative, yet accurate, approach for designing hardware specifications for nonlinear elements in all-digital mmWave massive MIMO. Scaling using a larger number of antennas with a smaller load factor is attractive, since the specifications for RF nonlinearities, baseband nonlinearities, and ADC precision can all be significantly relaxed by operating at lower load factors.  The requirements can also be relaxed by use of appropriate power control, as illustrated by the simple adaptive power control scheme considered here. 

While we have considered LoS channel models here, we note that our approach extends to sparse multipath channels.  At high symbol rates, equalization over a large delay spread becomes computationally unattractive. In this case paths that differ significantly in delay and angular spread from the dominant path play the role of additional interference, and can be folded into our framework.

In addition to the extensive effort required to realize our design prescriptions in hardware, there are also important open issues related to the digital backend, given the challenges of both computation and data transport for the multiGigabaud, multiuser system considered here.  Thus, despite the extensive prior research on multiuser detection, there are significant open issues on the design of strategies that are efficient enough (in terms of both computation and communication on the backend fabric) to scale with the number of antennas, number of users, and bandwidth. 
Preliminary results in \cite{abdelghany2019beamspace,abdelghany2019rainbow} indicate that exploiting channel sparsity is a promising approach for addressing such bottlenecks.

\appendices
\section{All-digital link budget} \label{Appendix0}

We provide here example parameters that demonstrate that the link budget for all-digital massive multiuser MIMO uplink system is realizable with low-cost silicon:
\begin{itemize}
\item antenna element gain covering a hemisphere is 3~dBi,
\item 16-element array at the mobile gives 12~dBi transmit beamforming gain, plus 12 dB power pooling gain,
\item 256-element array in the base station gives 24~dBi receive beamforming gain,
\item noise figure for each RF chain in the base station of 7~dB,
\item thermal noise power over 5~GHz bandwidth is about -77~dBm,
\item and free space path loss of an edge user at 100~m using a carrier frequency of 140~GHz is about 115~dB.
\end{itemize}
The transmit power required from each power amplifier (PA) at the mobile to achieve a target SNR (in dB) for an edge-user, namely $SNR_{edge}|_{dB}$, can now be computed as 
\begin{align}
P_{PA}=SNR_{edge}|_{dB}-9 ~~ \text{dBm}. \label{SNRLinkBuget} 
\end{align} 
For example, $SNR_{edge}|_{dB}$ of about 16~dB (shown to suffice for our case study) requires 7~dBm PA output, which is realizable in CMOS (CMOS designs of up to 11~dBm have
been reported in \cite{CMOSPA}).

\section{Uniform vs nonuniform quantization} \label{AppendixA}

\begin{figure}
\centering
\subfloat[]{ \includegraphics[trim=7.5cm 0 7.8cm 0,clip,scale=0.35]{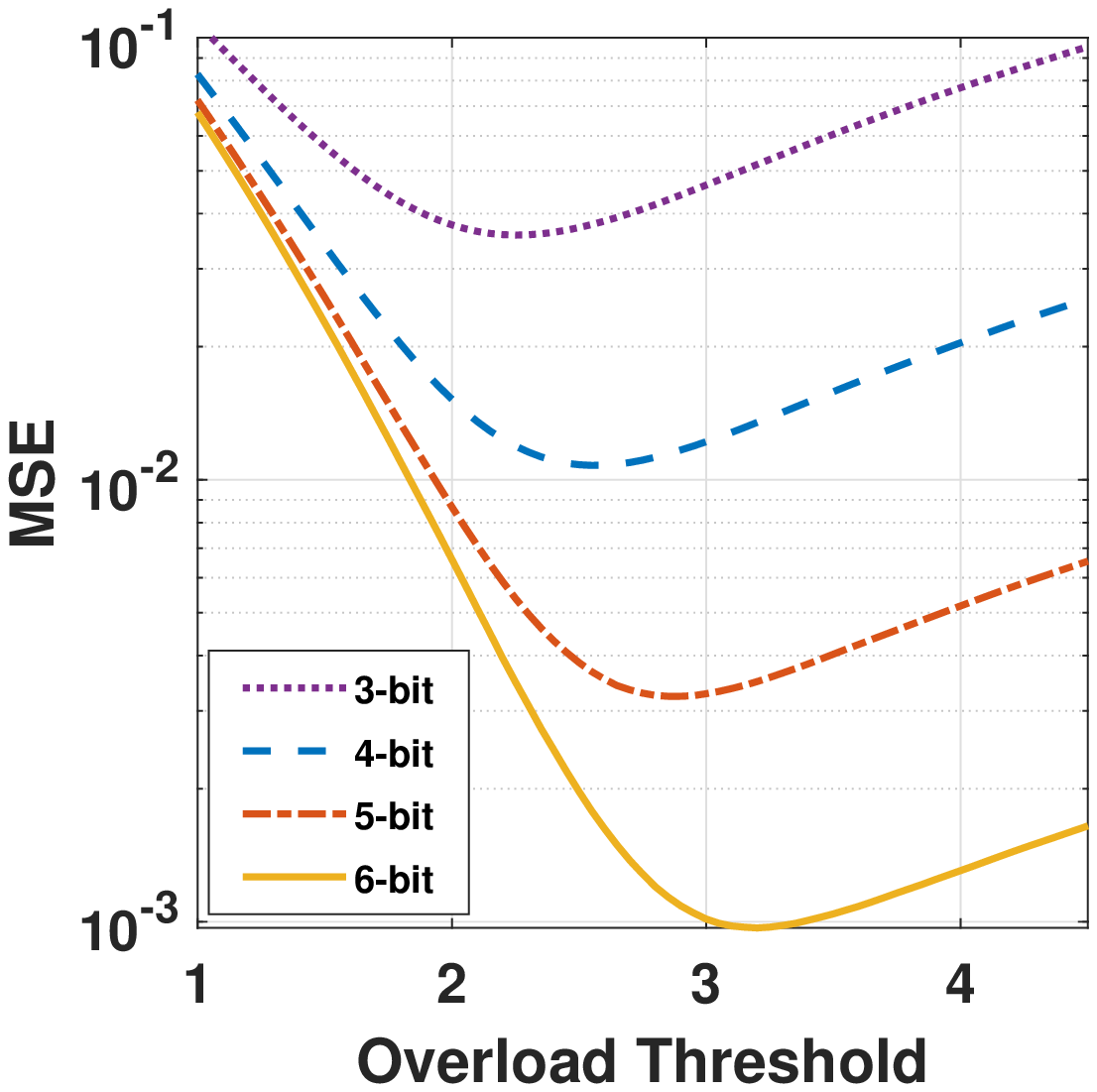}}
\subfloat[]{ \includegraphics[trim=7.5cm 0 7.8cm 0,clip,scale=0.35]{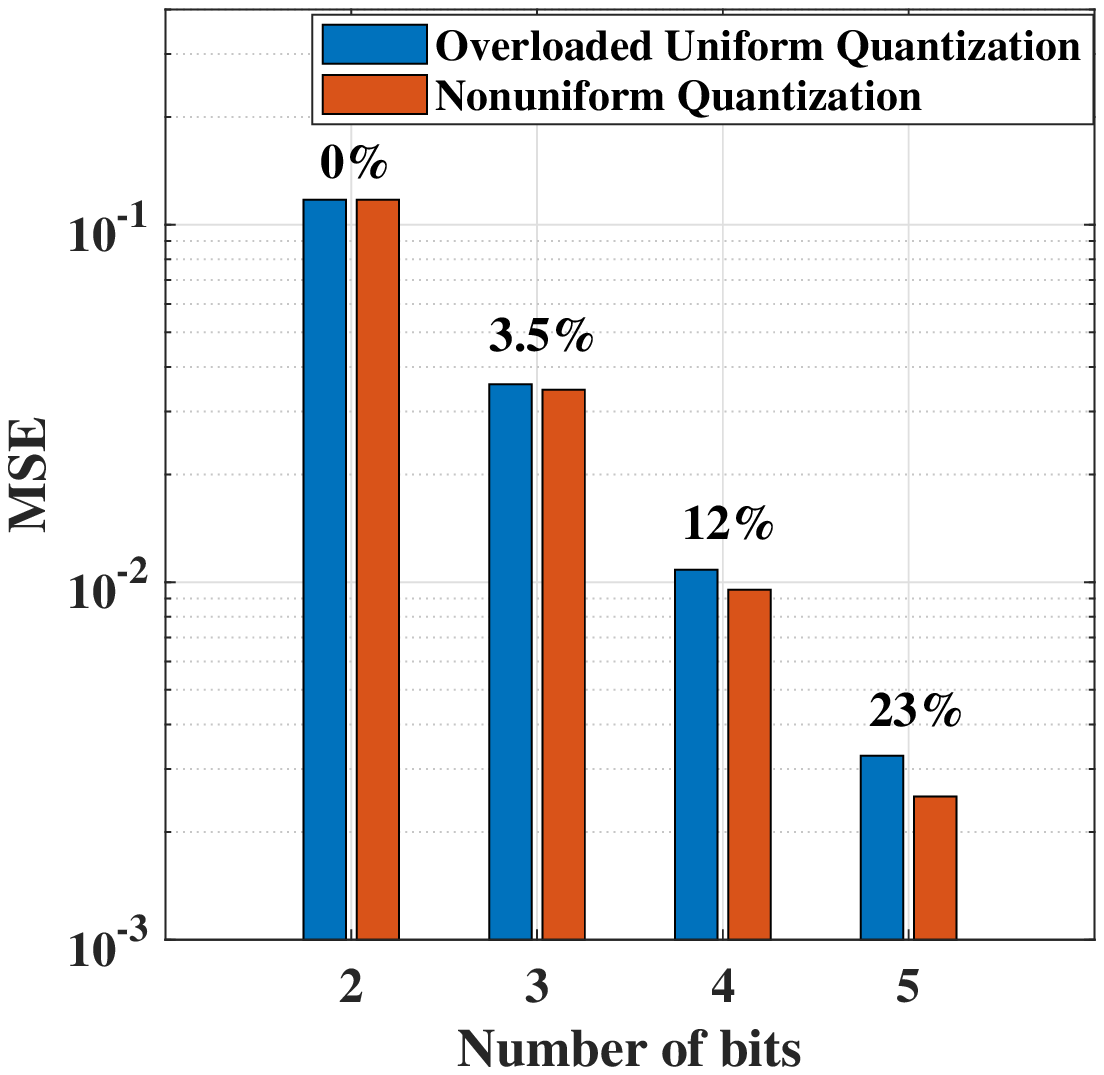}}

\caption{(a) MSE versus overload threshold. (b) MSE comparison of overload uniform quantizer versus MSE-optimal nonuniform quantizer. The percentages represent the
relative reduction in MSE from using MSE-optimal nonuniform quantization}
    \label{uniformVsNonuniform}
\end{figure}

Our simulation results are for an overloaded ADC. The overloaded uniform ADC comprises two regions in its I/O characteristic, the granular and overload regions. The granular region is quantized uniformly, with bounded quantization noise. While quantization noise in the overload region, represented by the quantizer levels
at the edges, is unbounded, the contribution to the MSE is kept comparable to that of the granular region by minimizing the MSE for the given input distribution;
see Fig. \ref{uniformVsNonuniform} (a), where MSE is plotted against overload threshold.

An alternative is to employ an MSE-optimal quantizer using Lloyd's algorithm \cite{lloyd1982least}, with quantization bins as listed in \cite{NonUniformQuantization}.
The MSE comparison between these two options is shown in Fig. \ref{uniformVsNonuniform} (b). The advantage of nonuniform MSE-optimal quantization is 
barely noticeable for the small number of quantization bits of interest here, hence we choose to work with the simpler overloaded uniform quantizer.

\section{Linear MMSE properties} \label{AppendixB}

From the point of view of a given user (the {\it desired} user) with channel $\mathbf{h}$, we may write the received signal corresponding to a single symbol as
\begin{equation} \label{LMMSE_model}
\mathbf{r} = b \mathbf{h} + \mathbf{w}_I + \mathbf{w}_N,
\end{equation}
where $b$ denotes the transmitted symbol, $\mathbf{w}_I$ denotes the interference vector and $\mathbf{w}_N \sim \mathcal{CN}( 0, \sigma_n^2 \mathbf{I} )$ denotes complex WGN.
Standard assumptions necessary for effective interference suppression are that the desired symbol is uncorrelated with the interference and noise:
$\mathbb{E}[b^* \mathbf{w}_I] = \mathbb{E}[b^* \mathbf{w}_N] = {\mathbf{0}}$.  We also assume that the interference and noise are uncorrelated.

A linear correlator $\mathbf{c}$ produces a decision statistic $\mathbf{c}^H \mathbf{r}$ for the desired symbol, and its SINR is given by
\begin{align} 
\nonumber
SINR ( \mathbf{c} ) =& \frac{ \mathbb{E}[ |b \mathbf{c}^H \mathbf{h} |^2 ] }{\mathbb{E}[| \mathbf{c}^H ( \mathbf{w}_I + \mathbf{w}_N)|^2]} \\
=& \frac{\sigma_b^2 | \mathbf{c}^H \mathbf{h} |^2}{\mathbf{c}^H \mathbf{R}_I \mathbf{c} + \sigma_n^2 || \mathbf{c} ||^2}, \label{SINR_defn}
\end{align}
where $\mathbf{R}_I = \mathbb{E}[ \mathbf{w}_I \mathbf{w}_I^H ]$ is the interference covariance matrix, and $\mathbf{R}_N = \mathbb{E}[\mathbf{w}_N \mathbf{w}_N^H ] = \sigma_n^2 \mathbf{I}$ is the noise covariance matrix.

The LMMSE correlator minimizes $MSE = \mathbb{E}[ |\mathbf{c}^H \mathbf{r} - b|^2]$ and maximizes SINR \cite{madhow1994mmse}.  For the additive noise-plus-interference model (\ref{LMMSE_model}), it
is known to be proportional to a whitened matched filter (i.e., it suppresses interference by whitening it):
\begin{align} 
\mathbf{c}_{MMSE} = \alpha ( \mathbf{R}_I + \mathbf{R}_N )^{-1} \mathbf{h} = \alpha ( \mathbf{R}_I + \sigma_n^2 \mathbf{I} )^{-1} \mathbf{h},
\end{align}
where $\alpha$ is a scale factor that can be solved for easily (e.g., see \cite{madhow1994mmse}). Since SINR does not depend on scale factor, it is easy to show, plugging into (\ref{SINR_defn}),
that
\begin{align} 
\nonumber
SINR =& \sigma_b^2 \mathbf{h}^H \left( \mathbf{R}_I + \mathbf{R}_N \right)^{-1} \mathbf{h}\\
 =& \sigma_b^2 \mathbf{h}^H  \left( \mathbf{R}_I + \sigma_n^2 \mathbf{I} \right)^{-1} \mathbf{h}. \label{SINR_max}
\end{align}

Let us also for reference define the SNR:
\begin{equation} \label{SNR_defn}
SNR = \sigma_b^2 \mathbf{h}^H \left( \mathbf{R}_N \right)^{-1} \mathbf{h} = \sigma_b^2 || \mathbf{h}||^2/ \sigma_n^2.
\end{equation}

\begin{rem}
A positive definite matrix $\mathbf{A} ( \theta )$ increases with $\theta$ if $\mathbf{A} ( \theta ) - \mathbf{A} (\theta') \geq {\mathbf{0}}$ for any
$\theta > \theta'$. That is, for any vector $\mathbf{u}$, $\mathbf{u}^H \mathbf{A} ( \theta ) \mathbf{u} \geq \mathbf{u}^H \mathbf{A} ( \theta' ) \mathbf{u}$.
\end{rem}
We can now infer the following properties relevant for our approach to performance analysis, stated as a lemma.
\noindent
\begin{lem}
{ If the noise level $\sigma_n^2$ increases, with the signal and interference characteristics unchanged, then\\
(a) Absolute performance gets worse, with $SINR$ and $SNR$ both decreasing.\\
(b) The noise enhancement gets better: $\frac{SNR}{SINR}$ decreases.}\\
\end{lem}
\noindent
\begin{proof}
For (a), we note that the positive definite matrix $\mathbf{R}_I + \sigma_n^2 \mathbf{I}$ increases with $\sigma_n^2$, hence its inverse decreases
with $\sigma_n^2$.  For (b), note that
\begin{align} 
\nonumber
\frac{SNR}{SINR} =& \frac{||\mathbf{h}||^2/ \sigma_n^2 }{\mathbf{h}^H  \left( \mathbf{R}_I + \sigma_n^2 \mathbf{I} \right)^{-1} \mathbf{h}}, \\   =& \frac{||\mathbf{h}||^2}{\mathbf{h}^H  \left( \mathbf{R}_I/\sigma_n^2 + \mathbf{I} \right)^{-1} \mathbf{h}}. \label{noise_enhancement}
\end{align}
The positive definite matrix  $\mathbf{R}_I/\sigma_n^2 + \mathbf{I}$ decreases with $\sigma_n^2$, hence its inverse increases with $\sigma_n^2$.
Thus, the denominator on the right-hand side of equation (\ref{noise_enhancement}) increases with $\sigma_n^2$, while the numerator is independent of it,
proving the desired result.\\
\end{proof}

\section*{Acknowledgment}

This work was supported in part by the Semiconductor Research Corporation (SRC) under the JUMP program (2018-JU-2778) and by DARPA (HR0011-18-3-0004). Use was made of the computational facilities administered by the Center for Scientific Computing at the CNSI and MRL (an NSF MRSEC; DMR-1720256) and purchased through NSF CNS-1725797.

\bibliographystyle{IEEEtran}
\bibliography{IEEEabrv,orgnizations//Ref}

\begin{thebibliography}{10}
\providecommand{\url}[1]{#1}
\csname url@samestyle\endcsname
\providecommand{\newblock}{\relax}
\providecommand{\bibinfo}[2]{#2}
\providecommand{\BIBentrySTDinterwordspacing}{\spaceskip=0pt\relax}
\providecommand{\BIBentryALTinterwordstretchfactor}{4}
\providecommand{\BIBentryALTinterwordspacing}{\spaceskip=\fontdimen2\font plus
\BIBentryALTinterwordstretchfactor\fontdimen3\font minus
  \fontdimen4\font\relax}
\providecommand{\BIBforeignlanguage}[2]{{%
\expandafter\ifx\csname l@#1\endcsname\relax
\typeout{** WARNING: IEEEtran.bst: No hyphenation pattern has been}%
\typeout{** loaded for the language `#1'. Using the pattern for}%
\typeout{** the default language instead.}%
\else
\language=\csname l@#1\endcsname
\fi
#2}}
\providecommand{\BIBdecl}{\relax}
\BIBdecl

\bibitem{yan2018performance}
H.~Yan, S.~Ramesh, T.~Gallagher, C.~Ling, and D.~Cabric, ``Performance, power,
  and area design trade-offs in millimeter-wave transmitter beamforming
  architectures,'' \emph{IEEE Circuits and Systems Magazine}, vol.~19, no.~2,
  2019.

\bibitem{razavi1998rf}
B.~Razavi and R.~Behzad, \emph{{RF} microelectronics}.\hskip 1em plus 0.5em
  minus 0.4em\relax Prentice Hall New Jersey, 1998, vol.~2.

\bibitem{bussgang}
J.~BUSSGANG, ``Crosscorrelation functions of amplitude-distorted gaussian
  signals,'' \emph{MIT Res. Lab. Elec. Tech. Rep.}, vol. 216, 1952.

\bibitem{bjornson2014massive}
E.~Bj{\"o}rnson, J.~Hoydis, M.~Kountouris, and M.~Debbah, ``Massive {MIMO}
  systems with non-ideal hardware: Energy efficiency, estimation, and capacity
  limits,'' \emph{IEEE Transactions on Information Theory}, vol.~60, no.~11,
  2014.

\bibitem{fan2015uplink}
L.~Fan, S.~Jin, C.-K. Wen, and H.~Zhang, ``Uplink achievable rate for massive
  {MIMO} systems with low-resolution {ADC},'' \emph{IEEE Communications
  Letters}, vol.~19, no.~12, 2015.

\bibitem{z2016spectral}
J.~Zhang, L.~Dai, S.~Sun, and Z.~Wang, ``On the spectral efficiency of massive
  {MIMO} systems with low-resolution {ADCs},'' \emph{IEEE Communications
  Letters}, vol.~20, no.~5, 2016.

\bibitem{xu2019uplink}
L.~Xu, X.~Lu, S.~Jin, F.~Gao, and Y.~Zhu, ``On the uplink achievable rate of
  massive {MIMO} system with low-resolution {ADC} and {RF} impairments,''
  \emph{IEEE Communications Letters}, vol.~23, no.~3, 2019.

\bibitem{mollen2017achievable}
C.~Moll{\'e}n, J.~Choi, E.~G. Larsson, and R.~W. Heath, ``Achievable uplink
  rates for massive {MIMO} with coarse quantization,'' in \emph{2017 IEEE
  International Conference on Acoustics, Speech and Signal Processing
  (ICASSP)}.\hskip 1em plus 0.5em minus 0.4em\relax IEEE, 2017.

\bibitem{jacobsson2017throughput}
S.~Jacobsson, G.~Durisi, M.~Coldrey, U.~Gustavsson, and C.~Studer, ``Throughput
  analysis of massive {MIMO} uplink with low-resolution {ADCs},'' \emph{IEEE
  Transactions on Wireless Communications}, vol.~16, no.~6, 2017.

\bibitem{studer2016quantized}
C.~Studer and G.~Durisi, ``Quantized massive {MU-MIMO-OFDM} uplink,''
  \emph{IEEE Transactions on Communications}, 2016.

\bibitem{j2018massive}
S.~Jacobsson, U.~Gustavsson, G.~Durisi, and C.~Studer, ``Massive {MU-MIMO-OFDM}
  uplink with hardware impairments: Modeling and analysis,'' in \emph{2018 52nd
  Asilomar Conference on Signals, Systems, and Computers}.\hskip 1em plus 0.5em
  minus 0.4em\relax IEEE, 2018.

\bibitem{LoS1}
A.~Maltsev, A.~Pudeyev, I.~Karls, I.~Bolotin, G.~Morozov, R.~Weiler, M.~Peter,
  and W.~Keusgen, ``Quasi-deterministic approach to {mmWave} channel modeling
  in a non-stationary environment,'' in \emph{2014 IEEE Globecom Workshops (GC
  Wkshps)}.\hskip 1em plus 0.5em minus 0.4em\relax IEEE, 2014.

\bibitem{LoS2}
T.~S. Rappaport, S.~Sun, R.~Mayzus, H.~Zhao, Y.~Azar, K.~Wang, G.~N. Wong,
  J.~K. Schulz, M.~Samimi, and F.~Gutierrez, ``Millimeter wave mobile
  communications for {5G} cellular: It will work!'' \emph{IEEE access}, vol.~1,
  2013.

\bibitem{LoS3}
T.~S. Rappaport, F.~Gutierrez, E.~Ben-Dor, J.~N. Murdock, Y.~Qiao, and J.~I.
  Tamir, ``Broadband millimeter-wave propagation measurements and models using
  adaptive-beam antennas for outdoor urban cellular communications,''
  \emph{IEEE transactions on antennas and propagation}, vol.~61, no.~4, 2012.

\bibitem{LoS4}
M.~Jacob, S.~Priebe, R.~Dickhoff, T.~Kleine-Ostmann, T.~Schrader, and
  T.~Kurner, ``Diffraction in mm and sub-mm wave indoor propagation channels,''
  \emph{IEEE Transactions on Microwave Theory and Techniques}, vol.~60, no.~3,
  2012.

\bibitem{AsilomarPaper}
M.~Abdelghany, A.~Farid, U.~Madhow, and M.~Rodwell, ``Towards all-digital
  {mmWave} massive {MIMO}: Designing around nonlinearities,'' in \emph{Asilomar
  Conference on Signals, Systems, and Computers}.\hskip 1em plus 0.5em minus
  0.4em\relax IEEE, 2018.

\bibitem{gersho2012vector}
A.~Gersho and R.~M. Gray, \emph{Vector quantization and signal
  compression}.\hskip 1em plus 0.5em minus 0.4em\relax Springer Science \&
  Business Media, 2012.

\bibitem{madhow1994mmse}
U.~Madhow and M.~L. Honig, ``{MMSE} interference suppression for
  direct-sequence spread-spectrum {CDMA},'' \emph{IEEE transactions on
  communications}, vol.~42, no.~12, 1994.

\bibitem{Verdu_book}
S.~Verdu \emph{et~al.}, \emph{Multiuser detection}.\hskip 1em plus 0.5em minus
  0.4em\relax Cambridge university press, 1998.

\bibitem{hajek2015random}
B.~Hajek, \emph{Random processes for engineers}.\hskip 1em plus 0.5em minus
  0.4em\relax Cambridge university press, 2015.

\bibitem{minkoff1985role}
J.~Minkoff, ``The role of {AM-to-PM} conversion in memoryless nonlinear
  systems,'' \emph{IEEE Transactions on Communications}, vol.~33, no.~2, 1985.

\bibitem{bj2018hardware}
E.~Bj{\"o}rnson, L.~Sanguinetti, and J.~Hoydis, ``Hardware distortion
  correlation has negligible impact on {UL} massive {MIMO} spectral
  efficiency,'' \emph{IEEE Transactions on Communications}, vol.~67, no.~2,
  2018.

\bibitem{jacobsson2017quantized}
S.~Jacobsson, G.~Durisi, M.~Coldrey, T.~Goldstein, and C.~Studer, ``Quantized
  precoding for massive {MU-MIMO},'' \emph{IEEE Transactions on
  Communications}, vol.~65, no.~11, 2017.

\bibitem{Balanis}
C.~A. Balanis, \emph{Antenna Theory: Analysis and Design}.\hskip 1em plus 0.5em
  minus 0.4em\relax New York, NY, USA: Wiley-Interscience, 2005.

\bibitem{UlukusYates}
S.~Ulukus and R.~D. Yates, ``Adaptive power control and {MMSE} interference
  suppression,'' \emph{Wireless Networks}, 1998.

\bibitem{abdelghany2019beamspace}
M.~Abdelghany, U.~Madhow, and A.~T{\"o}lli, ``Beamspace local {LMMSE}: An
  efficient digital backend for {mmWave} massive {MIMO},'' in \emph{2019 IEEE
  20th International Workshop on Signal Processing Advances in Wireless
  Communications ({SPAWC})}.\hskip 1em plus 0.5em minus 0.4em\relax IEEE, 2019,
  pp. 1--5.

\bibitem{abdelghany2019rainbow}
M.~Abdelghany, U.~Madhow, and M.~Rodwell, ``An efficient digital backend for
  wideband single-carrier {mmWave} massive {MIMO},'' in \emph{to be presented
  in IEEE Global Communications Conference (Globecom), Waikoloa, Hawaii, Dec.
  2019}.

\bibitem{CMOSPA}
D.~Simic and P.~Reynaert, ``A 14.8 {dBm} 20.3 {dB} power amplifier for {D-band}
  applications in 40 nm {CMOS},'' in \emph{2018 IEEE Radio Frequency Integrated
  Circuits Symposium (RFIC)}.\hskip 1em plus 0.5em minus 0.4em\relax IEEE,
  2018.

\bibitem{lloyd1982least}
S.~Lloyd, ``Least squares quantization in {PCM},'' \emph{IEEE transactions on
  information theory}, vol.~28, no.~2, 1982.

\bibitem{NonUniformQuantization}
J.~Max, ``Quantizing for minimum distortion,'' \emph{IRE Transactions on
  Information Theory}, vol.~6, no.~1, 1960.

\end{thebibliography}
\printnomenclature
\end{document}